\documentclass[]{scrartcl}
\usepackage{graphicx}
\usepackage{natbib}
\usepackage{subcaption}
\usepackage{booktabs}
\usepackage{amsmath}
\usepackage{hyperref}

\usepackage{dcolumn}
\usepackage{booktabs}
\usepackage{tikz}
\usetikzlibrary{positioning,shapes,arrows, fit}
\usepackage{pgfplots}
\pgfplotsset{compat=1.6}
\newcolumntype{M}[1]{D{.}{.}{1.#1}}
\usepackage{amsmath}

\usepackage{geometry}
\newgeometry{vmargin={25mm}, hmargin={12mm,12mm}}

\usepackage{amssymb}
\usepackage{pifont}
\newcommand{\cmark}{\ding{51}}
\newcommand{\xmark}{\ding{55}}

\usepackage{authblk}

\title{\vspace{-1.5cm} \noindent\Large\textbf{Emergency Response Inference Mapping (ERIMap): \\ 
A Bayesian Network-based Method for Dynamic Observation Processing \\ in Spatially Distributed Emergencies}}

\date{\vspace{-7ex}}

\small
\author[1, *]{\small Moritz Schneider}
\author[1]{\small Lukas Halekotte}
\author[2]{\small Tina Comes}
\author[1]{\small Daniel Lichte}
\author[3]{\small Frank Fiedrich}
\affil[1]{\small German Aerospace Center (DLR), Institute for the Protection of Terrestrial Infrastructures, Germany}
\affil[2]{Faculty of Technology, Policy and Management, TU Delft, The Netherlands}
\affil[3]{Chair for Public Safety and Emergency Management, University of Wuppertal, Germany} 
\affil[*]{Corresponding author, e-mail: moritz.schneider@dlr.de} 

\begin{document}
\maketitle

\par\noindent\rule{\textwidth}{0.3pt}
\begin{abstract}
In emergencies, high stake decisions often have to be made under time pressure and strain. In order to support such decisions, information from various sources needs to be collected and processed rapidly. The information available tends to be temporally and spatially variable, uncertain, and sometimes conflicting, leading to potential biases in decisions. Currently, there is a lack of systematic approaches for information processing and situation assessment which meet the particular demands of emergency situations. To address this gap, we present a Bayesian network-based method called \textit{ERIMap} that is tailored to the complex information-scape during emergencies. The method enables the systematic and rapid processing of heterogeneous and potentially uncertain observations and draws inferences about key variables of an emergency. It thereby reduces complexity and cognitive load for decision makers. The output of the \textit{ERIMap} method is a dynamically evolving and spatially resolved map of beliefs about key variables of an emergency that is updated each time a new observation becomes available. The method is illustrated in a case study in which an emergency response is triggered by an accident causing a gas leakage on a chemical plant site. 
\end{abstract}
\par\noindent\rule{\textwidth}{0.3pt}


\section{Introduction}
Situation awareness is crucial to emergency response. To improve the awareness and understanding of the situation, information from various sources is collected, assessed and combined \citep{Endsley1995, Javed2011}. In emergency response, the process of building situation awareness often has to be performed under considerable time pressure, even though the stakes are extremely high \citep{Comes2015, Abdalla2016, Hao2018}. What is more, emergencies usually are complex situations which are characterised by a multitude of spatially distributed and dynamically evolving factors. Accordingly, heterogeneous and uncertain information about very different aspects needs to be continuously combined \citep{Comes.2012} to understand the situation on the ground, and its implications for people and livelihoods. To be sure, analysing the situation is a continuous process that is interlaced with decision making \citep{comes2020}: as decision makers assess their options, new information becomes available to which plans need to be continuously adapted \citep{Comes.2012}. \\

In light of the complexity of emergency situations and the time pressure under which emergency responders usually operate, automated or semi-automated methods which are capable of efficiently processing the incoming information and mapping the key aspects of an emergency situation in a condensed manner are of great value for supporting decision makers in emergency response \citep{Avvenuti2018}. In order to do so, such methods must meet the specific properties of typical observations in an emergency. This is exactly what we set out to do with this work - i.e., we aim for a method which is tailored to the specific information-scape in emergency response. To this end, we first derive six requirements (R1-R6) which such a method should meet in order to allow for a versatile and comprehensive processing and mapping of information which become available in emergency situations.

\subsection{Information-Scape in Emergency Response}
In an emergency, the information available typically presents a fragmented description of the actual situation -- especially in the initial stages of an emergency, information about key variables is often scarce or, at least, incomplete \citep{Li2023, Wu2021}. A suitable method should therefore be capable of providing meaningful insights based on limited or incomplete information (\textbf{R1: }process \textit{incomplete} information). In this regard, a method which is capable of utilising information from diverse sources -- e.g., individuals \citep{Guo2023, Fathi2020}, sensors \citep{Milana2022, Huang2022}, or geographic information systems (GIS) \citep{Tzavella2018, Geiss2022} -- is of great value, since it increases the amount of information which can be incorporated in the assessment of the situation (\textbf{R2:} process information from \textit{diverse sources}). However, when incorporating information from diverse sources, it should be considered that not every piece of information is unambiguous and not every information source is 100\% reliable (people make mistakes and sensors malfunction). For the processing of information, this means that the level of uncertainty associated with different information sources must be taken into consideration \citep{Comes2012} (\textbf{R3:} process \textit{uncertain} information). Another aspect when incorporating information from multiple sources - which might differ with regard to their specific perspectives, biases or levels of expertise \citep{Walle2008} - is that different observations can contradict each other (source A says Yes, source B says No). A suitable method should therefore incorporate a scheme for handling this type of noisy information (\textbf{R4:} process \textit{conflicting} information). \\

A characteristic aspect of assessing an emergency situation is that it is a dynamical task which extends over the entire course of the emergency - during the assessment, the situation itself as well as the available information about it develop dynamically \citep{Turoff2004}. For a suitable method this implies that it should be capable of dynamically incorporating new information – i.e., as the actual situation evolves and new observations trickle in, the assessment of the situation should evolve accordingly (\textbf{R5: }process \textit{dynamic} information). A last crucial aspect of emergency situations is their spatial dimension \citep{Abdalla2016}. Since a comprehensive understanding of the geographic extent of an emergency and its impact at different locations is essential for an effective emergency management (e.g., for prioritising response efforts), a suitable method should allow for processing and mapping the spatially distributed information characterising an emergency event (\textbf{R6: }process \textit{spatial} information). \\

To summarise, a method for processing observations in emergency response should meet the following six requirements:

\begin{itemize}
    \item R1: process \textit{incomplete} information 
    \item R2: process information from \textit{diverse sources} 
    \item R3: process \textit{uncertain} information 
    \item R4: process \textit{conflicting} information 
    \item R5: process \textit{dynamic} information 
    \item R6: process \textit{spatial} information 
\end{itemize}

\subsection{Research Gap and Main Contribution}
A conceptual basis which is particularly suitable for the creation of a method which meets these requirements are Bayesian networks (BNs) \cite{pearl1985bayesian}. BNs are probabilistic graphical models that are composed of nodes representing variables which characterise the state of a system and edges which describe the probabilistic dependencies between these variables. Drawing inferences on the states of some nodes based on incomplete information regarding other nodes of the network is one of the core features of a BN (R1). Integrating multiple sources to inform the BN is feasible since BNs generally allow for the incorporation of different types of input data (R2). The consideration of uncertainties associated with different observation sources is possible via the use of uncertain evidence \citep{Mrad2015} (R3). Furthermore, the use of uncertain evidence also enables dealing with contradictory observations (R4). Displaying the dynamic progression of key variables of a situation given new observations can also be achieved using a BN (R5). And finally, by combining a BN with a geographic information system (GIS), the spatial dimension of an emergency can be considered \citep{Johnson2012} (R6). \\

In this paper, we present \textit{ERIMap} (Emergency Response Inference Mapping), a new method for supporting situation awareness that is designed to take into account the specific information-scape in emergency response. Since BNs are potentially capable of meeting the six requirements (R1-R6) that characterise the typical information-scape in emergency response, we selected a BN as the core of our method. While several Bayesian network-based approaches have been put forward that address some of the aforementioned requirements (see section \ref{background_5properties}), so far there is no method that fulfils all of them. This is precisely what \textit{ERIMap} has to offer. This means that the method supports situation awareness by mapping inferences drawn from processing incomplete, uncertain, and conflicting observations from diverse sources which evolve dynamically and are spatially distributed. It thus allows for a fast processing of large amounts of diverse observations, which is crucial regarding the time pressure in emergencies and the complexity of the corresponding situations. \\

In the remainder of this work, first, some background on BNs is provided. Special emphasis is placed on the introduction of uncertain evidence, the combination of a BN and a GIS, and relevant literature that deals with observation processing in BNs. Second, our \textit{ERIMap} method is introduced. Third, our method is illustrated by a case study developed with practitioners from a plant fire brigade of Henkel, a multinational marked-listed chemical company, headquartered in Germany. In the scenario of the case study, a chlorine gas tank leak causes a gas dispersion throughout the plant site. Fourth, the results of the case study are presented using multiple synthetic outcomes of the scenario that include different observation sequences. Finally, the proposed method is discussed and future work is outlined.

\section{Background}
\subsection{Bayesian Networks}
\begin{figure}[!h]
	\centering
	\includegraphics{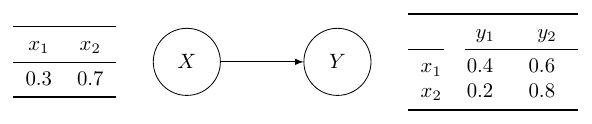}
	\caption{Example of a BN with two nodes, one edge, and respective marginal and conditional probability tables.}	\label{bn_ex}
\end{figure}

Bayesian networks (BNs) are probabilistic graphical models consisting of directed acyclic graphs \citep{pearl1985bayesian} (see Fig. \ref{bn_ex}). They present a powerful tool to embed knowledge and to perform belief updates about variables given new information about other variables. In particular, they allow to draw such inference on the basis of incomplete and uncertain evidence. Bayesian networks are used in a variety of research fields, such as diagnosis in medicine \citep{Seixas2014}, emergency response \citep{Ricci2024}, safety engineering \citep{RamirezAgudelo2021}, or decision making \citep{Waal2007}. For a comprehensive overview of topics, see \cite{Weber2012}. In the following, a brief introduction of the main components of BNs and relevant rules is outlined.

Bayesian networks are composed of nodes representing system variables as probability distributions and directed edges representing their probabilistic dependencies \citep{pearl1985bayesian}. Nodes can be either dependent (see node \textit{Y} in Fig. \ref{bn_ex}) or independent (see node \textit{X} in Fig. \ref{bn_ex}). An independent node is described by a Marginal Probability Table (MPT) (see left table in Fig. \ref{bn_ex}). A dependent node does have at least one parent node and is hence described as child node. To each dependent node, Conditional Probability Tables (CPT) are assigned (see right table in Fig. \ref{bn_ex}), containing one probability value for every possible combination of child node and parent node states. Given evidence on node $Y$, e.g. $Y$ is in state $y_{1}$, the Bayes' rule (see Eq. \ref{eq1}) can be applied to infer the probability of $X$ given new evidence, i.e. $P(X|Y=y_{1})$.

\begin{equation}\label{eq1}
	P(X|Y)=\frac{P(Y|X) \cdot P(X)}{P(Y)}
\end{equation}

\subsubsection{Evidence in Bayesian Networks} \label{evidence_chapter}
Belief updates in BNs require evidential findings (or observations) regarding the state of one or multiple nodes of the BN \citep{Pan2006}. Evidence in a BN can be either certain (also called hard evidence) or uncertain \citep{Mrad2012,YUN2010}. Uncertain evidence can be of two types: soft evidence \citep{Valtorta2002} or virtual evidence \citep{Pearl1988}. Each of the three types of evidence (hard, soft, and virtual) follows a different belief update rule \citep{YUN2010}.  \\

Given \textbf{hard evidence} on a node of a BN, the exact state of this node is known with certainty \citep{Mrad2012}. This means that an observation which provides hard evidence is considered to be undoubtedly true (in contrast to virtual evidence) and perfectly precise (in contrast to soft evidence). Entering hard evidence into a BN is straightforward: the respective node is simply set to the reported state (e.g., node $Y$ is in state $y_1$) or, in terms of likelihoods, the likelihood of the reported state is set to $1$ while the likelihood of all other states is set to $0$ (e.g., $L(Y)=(1,0)$). \\ 

\begin{figure}[!h]
	\centering
	\includegraphics{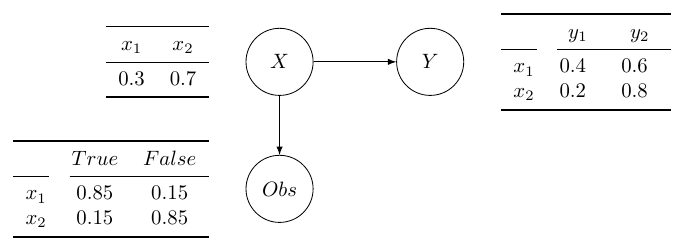}
	\caption{Illustrative example of a BN with two initial nodes and one virtual node.}	\label{bn_obs_example}
\end{figure}

\textbf{Virtual evidence} reflects uncertainty about whether a reported observation is true. It can thus be interpreted as \textit{evidence with uncertainty} \citep{YUN2010}, which is represented as a likelihood ratio \citep{Pearl1988}. Examples of virtual evidence are matters of varying veracity or accuracy, such as information provided by an imperfect sensor \citep{Mrad2015} or information provided by a person who has only partially observed an area \citep{BenMrad2013}. Given virtual evidence on a node of a BN, an additional virtual child node (node \textit{Obs} in Fig. \ref{bn_obs_example}) is attached to the respective node \citep{Mrad2012}. The initial CPT (likelihood ratio) of the binary child node represents the presumed likelihood that the underlying observation is true (85\% for node \textit{Obs} in Fig. \ref{bn_obs_example}). The belief update about the originally addressed node in the BN (node $X$ in Fig. \ref{bn_obs_example}) is then obtained by propagating hard evidence from the virtual child node, assuming that it is in state \textit{True} ($Obs$ = \textit{True}). 

\textbf{Soft evidence} considers the uncertainty which is included in a reported observation. It can thus be interpreted as \textit{evidence of uncertainty} \citep{YUN2010}, which can be represented as a probability distribution of one or more variables \citep{Valtorta2002}. Given soft evidence on a node of a BN, one is uncertain about the precise state of the node but certain about its probability distribution \citep{Mrad2012}. In contrast to virtual evidence, soft evidence can be interpreted as a new probability distribution of a variable that arose after creation of the model \citep{Mrad2015}. To enter soft evidence about one node into a BN, this evidence can be converted into a virtual evidence. To this end, the likelihood ratio of the additional virtual child node is calculated as the quotient of the probability ratio $\Lambda(X)$ and the prior probability of the addressed variable $P(X)$ (see Eq. \ref{eq3}). Subsequently, $L^{*}(X)$ is normalised to one (see Eq. \ref{eqnorm}). The following steps to perform belief updates in the BN are the same as for virtual evidence in case of a single soft evidence. 

\begin{equation} \label{eq3}
	L^{*}(X)=\dfrac{\Lambda(X)}{P(X)}
\end{equation} 
\begin{equation} \label{eqnorm}
	L(x_{1},...,x_{n})=\left(\dfrac{L(x_{1})}{\sum_{i=1}^{n}L(x_{i})},...,\dfrac{L(x_{n})}{\sum_{i=1}^{n}L(x_{i})}\right)
\end{equation}

The conversion from probability ratio into likelihood ratio compensates for the influence of the prior distribution of node $X$. Given the obtained virtual evidence with the likelihood ratio $L(X)$ on node $X$, the posterior probability of node $X$ is equal to the probability ratio $\Lambda(X)$ provided by the soft evidence. \\

Soft evidence can be fixed or not-fixed. Fixed soft evidence is implemented by assigning a new probability distribution for the respective variable and is considered as immutable, even in case of later observed evidence for other nodes in the BN \citep{Mrad2015}. In case of not-fixed soft evidence, the belief about the respective node can change in response to evidence for other nodes in the BN \citep{Mrad2015}. It should be noted that in this work we only consider not-fixed soft evidence and thus the term soft evidence always refers to not-fixed soft evidence.

\subsubsection{Bayesian Networks combined with Geographic Information Systems}
Bayesian networks are increasingly used for spatial inference. The interaction between a GIS and a BN can be bidirectional: GIS layers can be used as input for BN nodes and inference on BN nodes can be represented in a GIS. An example of a GIS input to and output from a BN is shown in \cite{Dlamini2011} who presented a BN model for fire risk mapping using GIS. Another example is shown in \cite{Wu2019} who developed a BN model with the goal of estimating the probability of a flood disaster. \\

To simplify and automate the link between the BN and the GIS, the attributes in the GIS layers must be linked to the corresponding states of the BN's variables. For example, a node \textit{Landuse} of a BN with states \textit{Forest}, \textit{Industrial}, and \textit{Urban} can be informed by a GIS layer that includes a spatial mapping of these three types of landuse. To create the output of the GIS-informed BN, the area under consideration must be divided into subset areas in the GIS, e.g. with a tessellation approach \citep{Rohr2020}. These subset areas determine the resolution of the subsequent analysis. A subset area should show attributes that are as homogeneous as possible. In each of these areas, inference in the BN is performed using the layer attributes of the respective area in the GIS. The results of the inference in the BN for a key variable, such as \textit{Risk of Fire} in \cite{Dlamini2011}, can be displayed using a heat map that colours the respective areas, depending on the probability of the risk for the respective area (e.g. see \cite{Wu2019, Dlamini2022}).

\subsection{Bayesian Networks for Observation Processing} \label{background_5properties}
The method developed in this paper aims at processing observations in a BN that are \textit{incomplete}, \textit{uncertain}, \textit{conflicting}, \textit{dynamic}, \textit{spatially distributed}, and from \textit{diverse sources}. Various authors addressed subsets of these potential properties of observations (see Table \ref{overview}). \\

\begin{table}[!h]
    \centering
    \caption{Summary of main references in regard to six properties of an observation.}
    \label{overview}
    \begin{tabular}{c|c|c|c|c|c|c}
    \toprule
        Reference & Incomplete & Diverse Sources & Uncertain & Conflicting & Dynamic & Spatial \\ \midrule
        \cite{Chan2005} & \cmark & \textcolor{gray}{\xmark} & \cmark & \textcolor{gray}{\xmark} & \textcolor{gray}{\xmark} & \textcolor{gray}{\xmark}  \\
        \cite{BenMrad2013} & \cmark & \cmark & \cmark & \textcolor{gray}{\xmark} & \textcolor{gray}{\xmark} & \textcolor{gray}{\xmark} \\
        \cite{Giordano2013} & \cmark & \textcolor{gray}{\xmark} & \cmark & \textcolor{gray}{\xmark} & \textcolor{gray}{\xmark} & \cmark \\
        \cite{Wu2019} & \cmark & \textcolor{gray}{\xmark} & \textcolor{gray}{\xmark} & \textcolor{gray}{\xmark} & \textcolor{gray}{\xmark} & \cmark \\
        \cite{Johnson2012} & \cmark & \textcolor{gray}{\xmark} & \textcolor{gray}{\xmark} & \textcolor{gray}{\xmark} & \textcolor{gray}{\xmark} & \cmark \\
        \cite{YUN2010} & \cmark & \textcolor{gray}{\xmark} & \cmark & \cmark & \textcolor{gray}{\xmark} & \textcolor{gray}{\xmark} \\
        \cite{Radianti2015a} & \cmark & \textcolor{gray}{\xmark} & \textcolor{gray}{\xmark} & \textcolor{gray}{\xmark} & \cmark & \cmark \\ 
        \cite{Valtorta2002} & \cmark & \cmark & \cmark & \textcolor{gray}{\xmark} & \textcolor{gray}{\xmark} & \textcolor{gray}{\xmark} \\ \midrule
      \textbf{ERIMap method} & \cmark & \cmark & \cmark & \cmark & \cmark & \cmark \\ 
\bottomrule
    \end{tabular}
\end{table}

We refer here to the field of BNs in general, not only to applications for emergency management. \cite{Chan2005} dealt with the question on how to capture informal statements as uncertain evidence in a BN. \cite{BenMrad2013} introduced multiple examples of uncertain evidence that include multiple potential observation sources. \cite{Giordano2013} used a BN to support conflict analysis for groundwater protection, utilising a GIS that provides soft evidence for each considered geographical area. \cite{Wu2019} connected a BN and a GIS to enable spatial analysis of flood disaster risk. \cite{Johnson2012} provided an overview of potential connections of a BN and a GIS. \cite{YUN2010} introduced multiple examples of virtual and soft evidence including consideration of observation (time) sequence and inconsistency (i.e. conflicting) observations. \cite{Radianti2015a} describe a spatio-temporal model based on a dynamic BN with the intent of supporting real-time evacuation planning. \cite{Valtorta2002} introduced soft evidence in general and outlined multiple examples of uncertain evidence that include different observation sources. \\

Table \ref{overview} clearly shows that no reference addresses all six requirements for observation processing in emergency response. To the best of our knowledge, no method has yet been presented that addresses all aspects. In this paper, we propose a method that covers all six requirements for information processing in emergency response. 

\section{ERIMap Method: Emergency Response Inference Mapping}
\begin{figure}[!h]
	\centering
	\includegraphics[width=16.5cm]{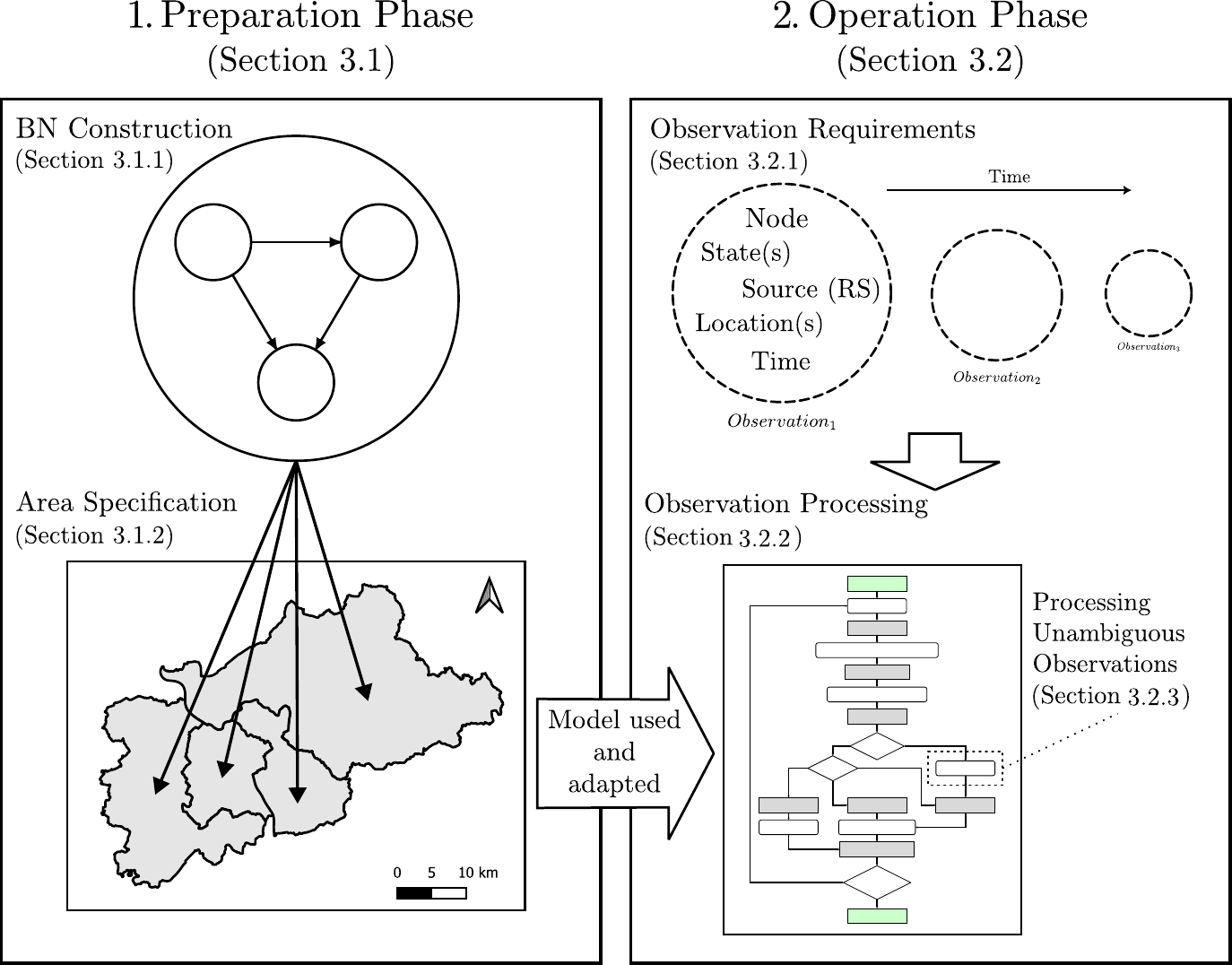}
	\caption{Overview of the structure of the method and division of the following sections. The \textit{preparation phase} includes the construction of the BN and specification of areas or locations. During the emergency \textit{operation phase}, observations are collected and processed.}	\label{summary}
\end{figure}

The goal of the \textit{ERIMap} method, is to draw inferences by processing observations from \textit{multiple sources}, which may be \textit{incomplete}, \textit{uncertain}, \textit{conflicting}, \textit{dynamic} and \textit{spatially distributed}. The application of the method is divided into two phases: the \textit{preparation phase} which takes place before an event (see left side of Fig. \ref{summary}) and the \textit{operation phase} which describes the application of the method during an emergency (see right side of Fig. \ref{summary}).  

\subsection{Preparation Phase}
\subsubsection{Bayesian Network Construction}
The first step of the preparation phase is the construction of a BN model for the intended area of application. The BN should include all variables that are key for decision making in a specific emergency (e.g. a flood or a forest fire scenario) as well as variables that directly or indirectly influence (the belief about) these key variables \citep{Wu2019}. To make sure that the \textit{ERIMap} method meets the demands of the users, all considered variables and the relationships between them should be identified in cooperation with decision makers in emergency response. Furthermore, additional sources can be incorporated to determine the probability tables of the BN (MPT and CPT), e.g., historical data or expert knowledge \citep{Druzdzel2000}.  

\subsubsection{Area Specification}
\begin{figure}[!h]
	\centering
	\includegraphics[width=8cm]{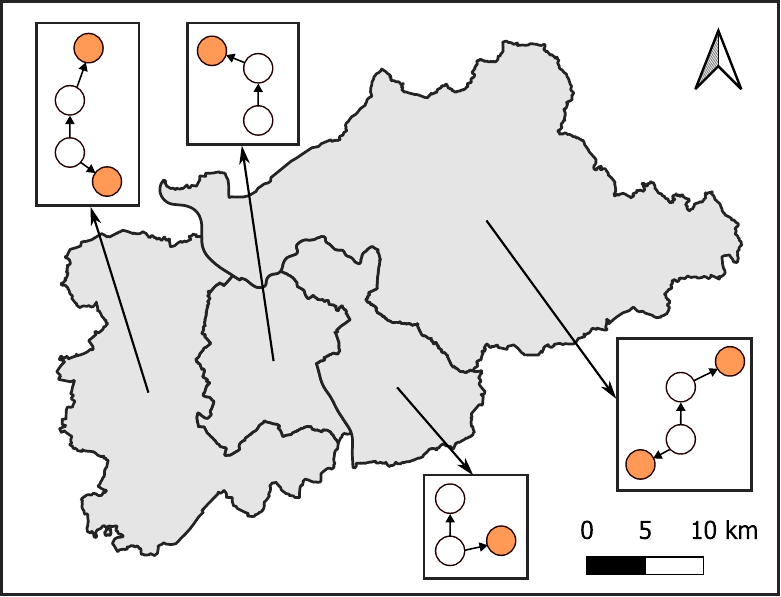}
	\caption{Illustration of one initial BN (white filled nodes) duplicated for four different areas. For each area, different virtual nodes (orange filled) are added.}	\label{areas_bn}
\end{figure}

In the second step of the preparation phase, the spatial resolution for the emergency consideration is specified - i.e., areas are specified which are to be assessed individually (see Fig. \ref{areas_bn}). Depending on the case, these areas can, for instance, correspond to districts, buildings, or specific point locations. To allow independent inference in the BN for each area, a duplicate of the initially constructed BN is assigned to each area (white nodes in Fig. \ref{areas_bn}). These initially identical BNs start to diverge as soon as they are fed with area-specific evidence - a process which is particularly impressive for uncertain evidence: Given uncertain evidence for a specific area, a virtual child node is added to the respective BN (orange nodes in Fig. \ref{areas_bn}); while BNs in other areas remain unchanged. Layers in the GIS that should serve as observation sources for the BN have to be linked to the attributes of the respective BN node states, i.e. they are used as inputs for BN nodes \citep{Johnson2012}. Besides using the GIS to inform the BN, the GIS serves to spatially display the results obtained from the BNs. \\

\subsection{Operation Phase}
\subsubsection{Observation Requirements} \label{info_req}
One of the core features of the proposed method is a procedure for translating different types of observations into evidence that can be considered in the BN. A necessary requirement for this transfer is that an observation contains five pieces of information:\\

(1) The \textbf{time} at which the observation has been conducted is used to display the temporal progression of the belief about variables in the BN. For example, it is important to know if an observation stating that people are in a building has been conducted before or after an evacuation of that building.  \\

(2) The \textbf{location} of the observation must be specified to enable the assignment of the observation to the respective area-specific BN(s). \\

(3) The \textbf{node} about which the observation provides evidence. This information is required to assign the observation to the respective node in the BN. \\

(4) The \textbf{reliability of the observation source}. A source can be fully reliable or not. The classification of reliability is important to avoid excessive influence of observations from unreliable sources. \\

(5) The \textbf{observed state(s)} of the node. Corresponding information can originate from (I) an unambiguous statement (e.g. "there is fire") or (II) an uncertain statement (e.g. "I think I saw a fire").  \\ 

\begin{addmargin}[25pt]{0pt}
For (I) - an unambiguous statement - hard or uncertain evidence can be considered, depending on the presumed reliability of the information source. For instance, a unambiguous observation (e.g., "there are no people in building A") provided by an emergency response team may be considered as 'confirmed' (hard evidence) while social media reports may be considered as 'not fully reliable' (uncertain evidence). In the latter case, the method introduced in Section \ref{single-state} is applied. \\
\end{addmargin}

\begin{addmargin}[25pt]{0pt}
For (II) - an uncertain statement - virtual evidence needs to be distinguished from soft evidence. In case of virtual evidence, the corresponding likelihood ratio is integrated by adding an additional virtual child node which influences the state of the respective parent node (see Section \ref{evidence_chapter}). For example, given a node called \textit{Fire}, which is used to infer the probability of a fire occurring, a potential indication child node could be a node \textit{Smoke Alarm} that is linked to a smoke detector. The likelihood values for false-positive and false-negative observations of the smoke detector constitute the likelihood ratio of the node \textit{Smoke Alarm}. By performing hard evidence on node \textit{Smoke Alarm}, $P(Fire|Smoke Alarm)$ is inferred using Eq. \ref{eq1}. In case of soft evidence, the obtained probability ratio describes a new probability distribution of a particular node and thus replaces the prior probability distribution following the routine outlined in Section \ref{evidence_chapter}. An example is a node \textit{Building Use} that has initially been set up and trained for a whole city in which 80\% of the buildings are apartment buildings and 20 \% are used for commercial purposes. If the same BN would now be used for a district in this city where commercial use of buildings is much more probable (e.g. 90\%), this new probability ratio shows a higher accuracy than the prior probability of the node \textit{Building Use} and is thus implemented using soft evidence. \\
\end{addmargin}

\subsubsection{Observation Processing}
\begin{figure}[!h]
	\centering
	\includegraphics[width=12cm]{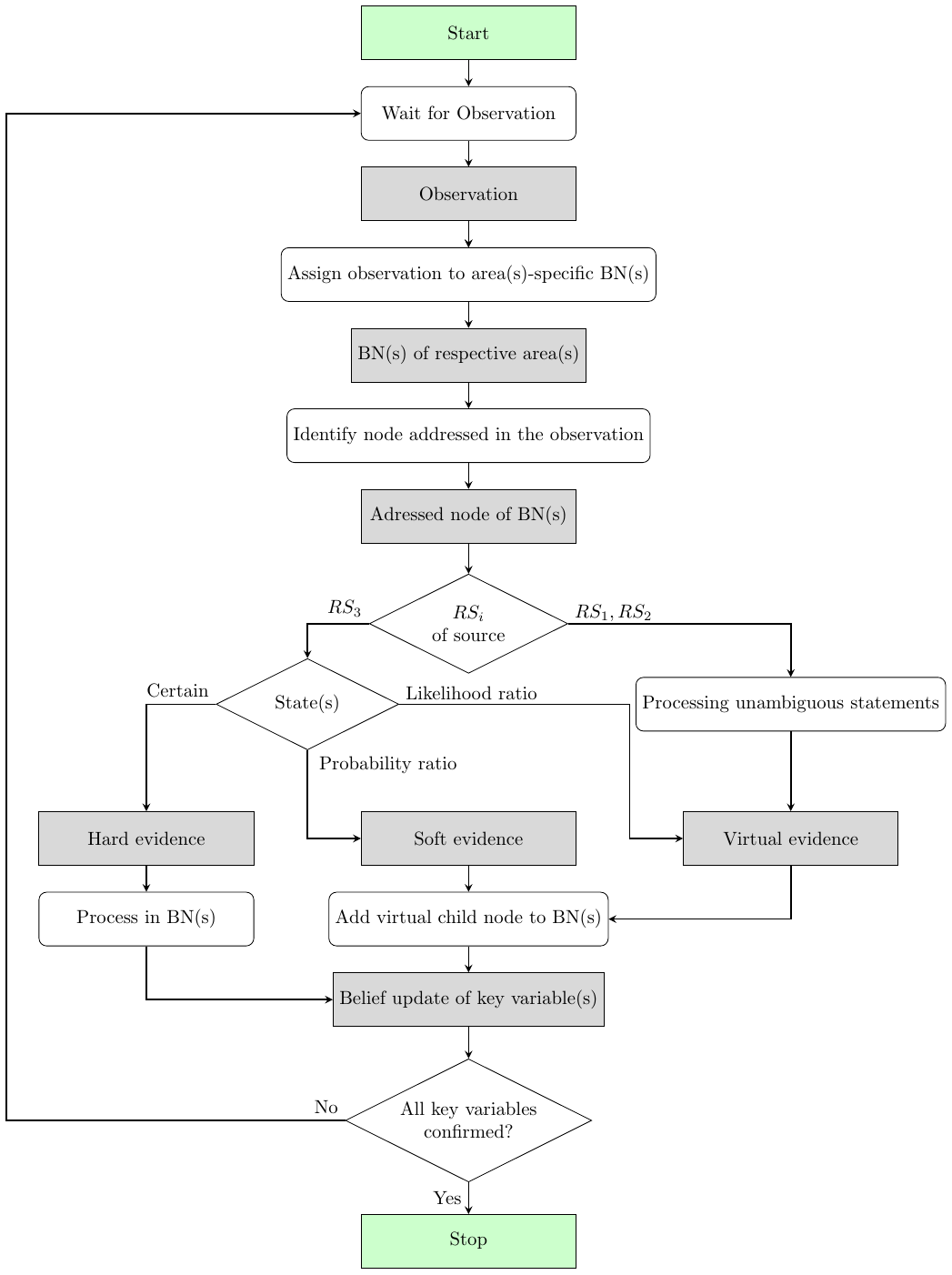}
	\caption{Summary of the \textit{ERIMap} method in the operation phase. Rectangles with rounded corners describe processes, grey rectangles describe the class of information. Decision nodes are diamond-shaped. Start and stop of the process are highlighted with green fill.}	\label{flowchart}
\end{figure}

In the operation phase, new observations are processed and fed into the area-specifc BNs according to a specific workflow (see Fig. \ref{flowchart}). Given a new observation, it is first assigned to the respective \textit{area}-specific BN(s) and then to the \textit{node} which is addressed in this observation. Afterwards, the method classifies the type of evidence (hard, soft, or virtual) based on the \textit{reliability} score of the observation source and on the reported \textit{node state(s)}. For an unambiguous observation provided by a source that shows a small or medium reliability ($RS_{1}$ and $RS_{2}$), the method introduced in Section \ref{single-state} is applied and a virtual child node is added to the respective node in the BNs. In case of an observation provided by a source that shows a high reliability ($RS_{3}$), three cases are distinguished: (1) an unambiguous observation without uncertainty that results in computing hard evidence, (2) a probability ratio, which shows a higher accuracy than the prior distribution and thus replaces it (soft evidence), and (3) a likelihood ratio that provides evidence with uncertainty (virtual evidence). In case (2) and (3) a virtual child node is added (see Section \ref{evidence_chapter}). Subsequently to this classification, a belief update for all key variables in the BNs of the respective areas is performed. The operation phase stops when all key variables are confirmed. \\ 

\subsubsection{Processing Unambiguous Observations} \label{single-state}
To account for uncertainty related to observations from unreliable sources, these observations are translated into virtual evidence \citep{Vomlel2004}. An observation source can be classified as unreliable for several reasons, e.g. (1) it is not known whether the person who provides an observation had access to all areas; (2) a person is not sure about an observation; (3) unreliable sensors. \\

To translate an unambiguous observation from an unreliable source into virtual evidence, two pieces of information are used in this method: the reliability score (RS) of the observation source and the criticality of the reported node state. For the RS, using predefined scores supports a quick classification during an emergency situation. We therefore define three example RSs which describe different degrees of reliability:\\

\begin{addmargin}[25pt]{0pt}
\begin{tabular}{ll}
    $RS_{1}$: & Small Reliability \\
    $RS_{2}$: & Medium Reliability  \\ 
    $RS_{3}$: & High Reliability  \\ 
\end{tabular} \\
\end{addmargin}

A likelihood is assigned to each $RS_{i}$ that quantifies the certainty of the observation. This likelihood, which can be interpreted as the chance that the observation is correct \citep{Chan2005}, is used to fill in the CPT of the respective virtual child node. Note that in case of a BN composed of only binary nodes, virtual evidence with a likelihood ratio of (0.5, 0.5) will show no effect on the respective posterior probability of the node. The likelihood values for the respective RSs should be selected in collaboration with potential users to reflect their preferences. The general case of a node $X$ with $N$ states, i.e. $X=\{x_{1},x_{2},...,x_{N}\}$, the likelihood ratio given an unambiguous observation stating node $X$ is in state $x_{1}$ is: $$L(X)=\left( L(RS_{i}), \dfrac{1-L(RS_{i})}{N-1},...,\dfrac{1-L(RS_{i})}{N-1} \right)$$
In this way, the posterior probability of state $x_{1}$ is increased (when performing hard evidence on the respective virtual child node) and at the same time the posterior probabilities of the other node states are decreased, while the ratio between the other node states remains the same. \\

In a next step, a regret function is introduced to better deal with conflicting observations. Using the example of a node \textit{People in Building}, one observation could state that people are in the building while another observation could state the opposite. In order to avoid that the two observations cancel each other out (assuming both sources share the same RS), the precautionary principle is applied: emphasise is placed on the node state that is more critical (i.e. people are in the building). This is achieved by increasing the likelihood of the critical node state by a certain percentage $\Theta$. The derived generalised likelihood ratio thus becomes: $$L^{*}(X)=\left( (L(RS_{i})+\Theta), \dfrac{1-(L(RS_{i})+\Theta)}{N-1},...,\dfrac{1-(L(RS_{i})+\Theta)}{N-1} \right)\;,$$\\
for $x_{1}$ being the observed critical node state. Note that the regret function is only applied to nodes whose states exhibit different levels of criticality. \\

\section{Case Study Preparation Phase}
\begin{figure}[!h]
	\centering
	\includegraphics[width=14cm]{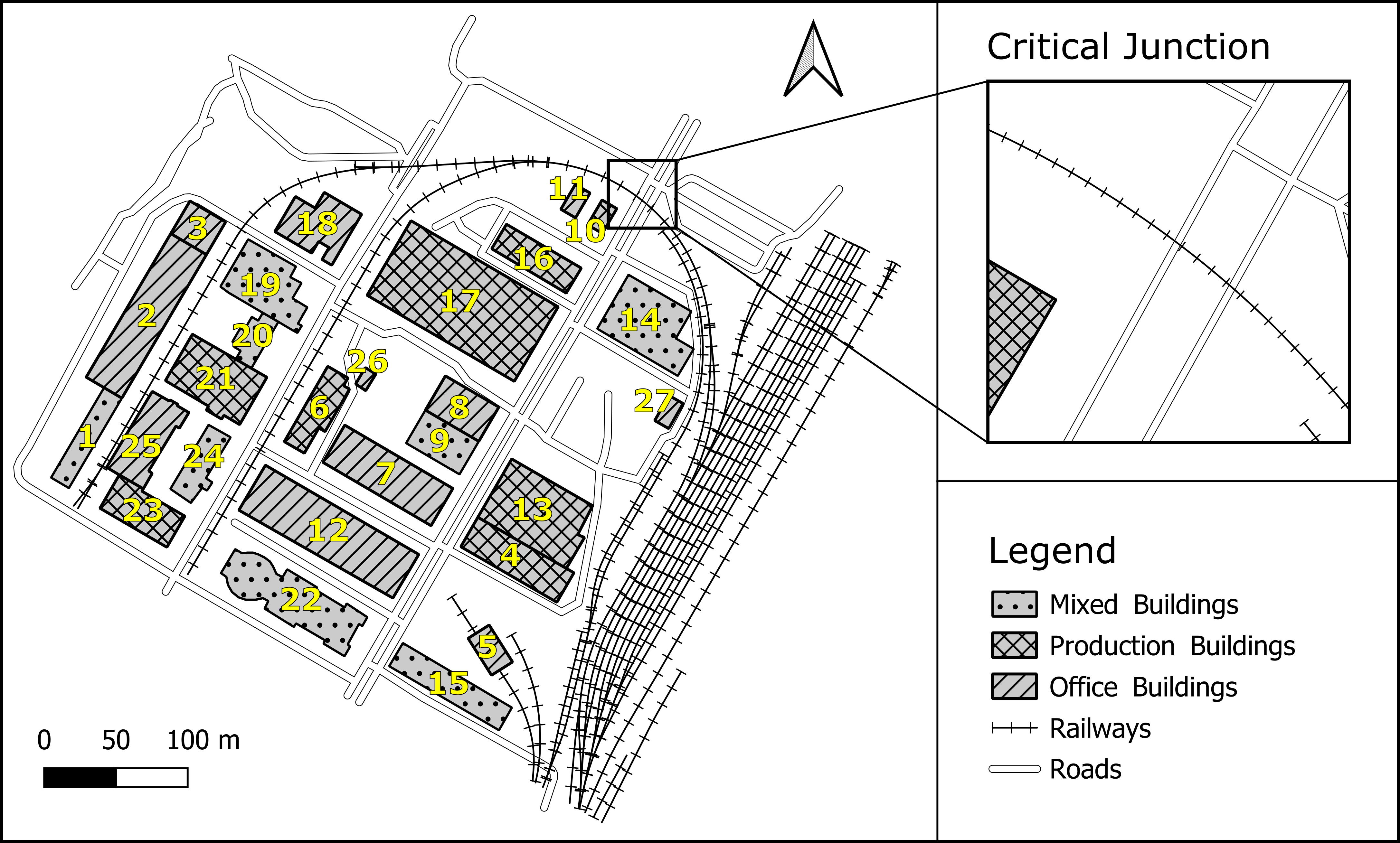}
	\caption{Map of the chemical plant. The map shows building types, building numbers, roads, railways and the critical junction. The chemical plant is located on a greenfield site, i.e. there are no surrounding buildings. }\label{case_study}
\end{figure}

In this section, our \textit{ERIMap} method is applied to a case study which has been developed in cooperation with the plant fire brigade of the German chemical company Henkel. A chemical plant site inspired by one of Henkel's sites is used as geographical setup (Fig. \ref{case_study}). The scenario is triggered by an accident between a truck and a tank wagon on a railway at a junction on the northern edge of the site (see top right of Fig. \ref{case_study}). The accident results in a gas leak, and potentially dangerous gas is dispersed throughout the site. In such an emergency, various sources of observations are to be expected. Geographic information systems, for example, support the simulation of gas dispersion, sensors are used to detect critical gas doses, and emergency responders inspect the buildings. Thus, in this case study, a dynamically evolving and spatially heterogeneous emergency situation whose evaluation can benefit from diverse observation sources is considered.  \\

The application of our method is designed to support the situation awareness of the plant fire brigade by helping them assess the risk that affected people are in a particular building. To this end, each building of the plant site is considered as an area to be assessed individually, i.e., each building receives a separate BN. Three types of building use are considered and randomly assigned (see Fig. \ref{case_study}). The case study is implemented in Python based on the libraries pgmpy \citep{ankan2015pgmpy} for Bayesian networks and GeoPandas \citep{kelsey_jordahl_2020_3946761} for geospatial data manipulation.

\subsection{Bayesian Network and Observation Sources}
\begin{figure}[!h]
	\centering
	\includegraphics[width=17cm]{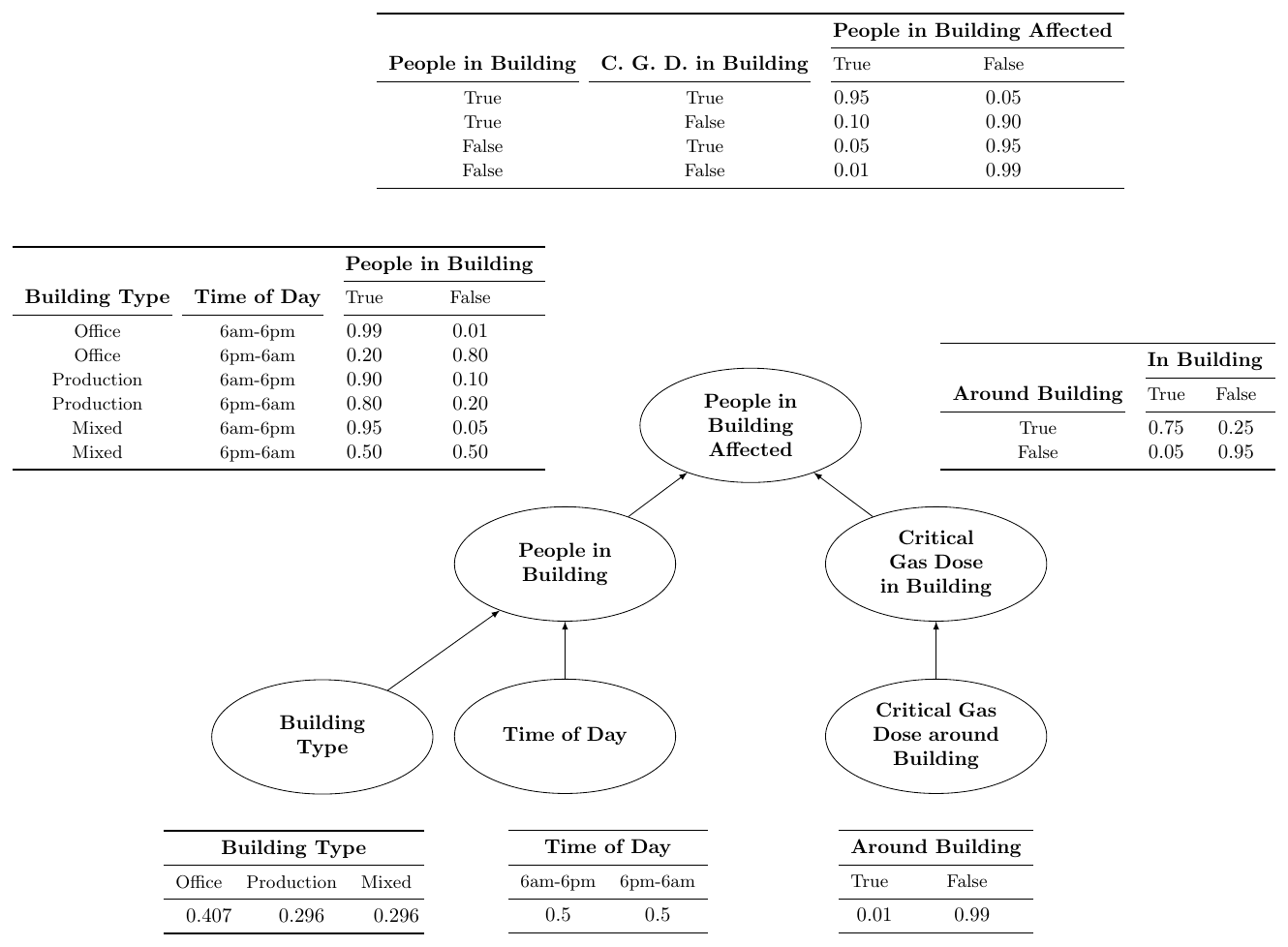}
	\caption{Bayesian network of the case study including seven variables, five edges, and the corresponding marginal and conditional probability tables.}	\label{bn_model}
\end{figure}

The BN of the case study is composed of six variables and five edges representing their probabilistic dependencies (see Fig. \ref{bn_model}). The target node of the BN is the variable \textit{People in Building Affected}, since this node is crucial for decision making and allocating rescue teams to buildings. Information about the presence of people in a building as well as the probability of a critical gas dose inside the building represent the parent nodes of \textit{People in Building Affected}. An additional node (\textit{Critical Gas Dose around Building}) is the parent node of \textit{Critical Gas Dose in Building}. This parent node is introduced to account for the uncertainty of gas dispersion from the surroundings of a building into the building itself. The presence of \textit{People in Building} can be inferred by its two parent nodes \textit{Building Type} and \textit{Time of Day}. Three building types are distinguished: office buildings (11 buildings) , production buildings (8 buildings), and mixed use buildings (8 buildings). \textit{Time of Day} shows two states: 6am - 6pm (day shift) and 6pm - 6am (night shift). In this case study, it is assumed that the presence of people in an office building during the night shift is less probable than in a production building. In a mixed-use building, the probability of human presence in the building is between that of office and production buildings. During the day shift, the probability of people being in a building is high for all building types. All other nodes are binary with state names \textit{True} and \textit{False}. The probability values used to fill in the MPT and CPT of the BN are selected by the authors.\\

Table \ref{sources} shows all considered information sources including the node(s) about which each source can provide observations and the respective reliability score(s). Additionally, it is stated whether the source provides an unambiguous observation or several uncertain states. Gas sensors and the simulation in GIS are the sources that do not provide unambiguous observations. The gas sensor is assumed to operate with a known accuracy. Therefore, this source is classified with $RS_{3}$, but does not provide an exact statement - it provides virtual evidence (see Fig. \ref{flowchart}). For the simulation of the gas dispersion, the prior probability of node \textit{Critical Gas Dose around Building} is $(True=0.01, False=0.99)$ due to the fact that a critical gas dose is not expected without further indication. Given a simulation of the critical gas dose around the buildings (see Section \ref{simulation}), this observation provides a new probability distribution that shows a higher value than the prior probability and is thus considered as soft evidence. The other observation sources provide unambiguous statements but are of different reliability scores. The likelihood values for the RSs used in this case study are: 70\% certainty for $RS_{1}$, 80\% for $RS_{2}$, and $RS_{3}$ is considered as hard evidence (100\% certainty). Thus, for a binary node $V$ and an observation of $RS_{2}$ stating $V$ is in state $v_{1}$, the corresponding likelihood ratio for the CPT of the virtual child node of $V$ is $(0.8,0.2)$. The $\Theta$ value used for the regret function is assumed to be $10\%$.

\begin{table}[!h]
    \centering
    \caption{Information on the observation sources considered in the case study.}
    \label{sources}
    \begin{tabular}{c|c|c|c|c}
    \toprule
        Source Class & Source Name & Node(s) & State(s) & $RS_{i}$ \\ \midrule
        Person & Emergency Responder & \textit{People in Building}, \textit{People Affected} & unambiguous  & $RS_{3}$\\
               & Civilian & \textit{People in Building}, \textit{People Affected} & unambiguous &  $RS_{1}$,$RS_{2}$\\ \midrule
        Sensor & Clock & \textit{Time of Day} & unambiguous & $RS_{3}$\\
               & Gas Sensor & \textit{Critical Gas Dose in Building} & uncertain & $RS_{3}$\\ \midrule
        GIS    & Simulation & \textit{Critical Gas Dose around Building} & uncertain & $RS_{3}$\\
               & Buildings Layer & \textit{Building Type} & unambiguous & $RS_{3}$ \\ \bottomrule
    \end{tabular}
\end{table}

\subsection{Gas Dispersion Hazard}\label{simulation}
The gas dispersion caused by a leakage in the tank wagon carrying chlorine is simulated using the \textit{Areal Location of Hazardous Atmosphere} (ALOHA) software, a widely used tool for chemical emergencies. ALOHA provides a simplified but quick steady-state simulation of a gas dispersion of various chemicals under surrounding conditions using a Gaussian plume model \citep{Brusca2016}. The software provides three threat zones that are characterised by a steady-state gas concentration in these areas. These zones represent an equilibrium of gas concentrations in the atmosphere given constant surrounding conditions and gas leakage. \\

\begin{figure}[!h]
	\centering
	\includegraphics[width=12cm]{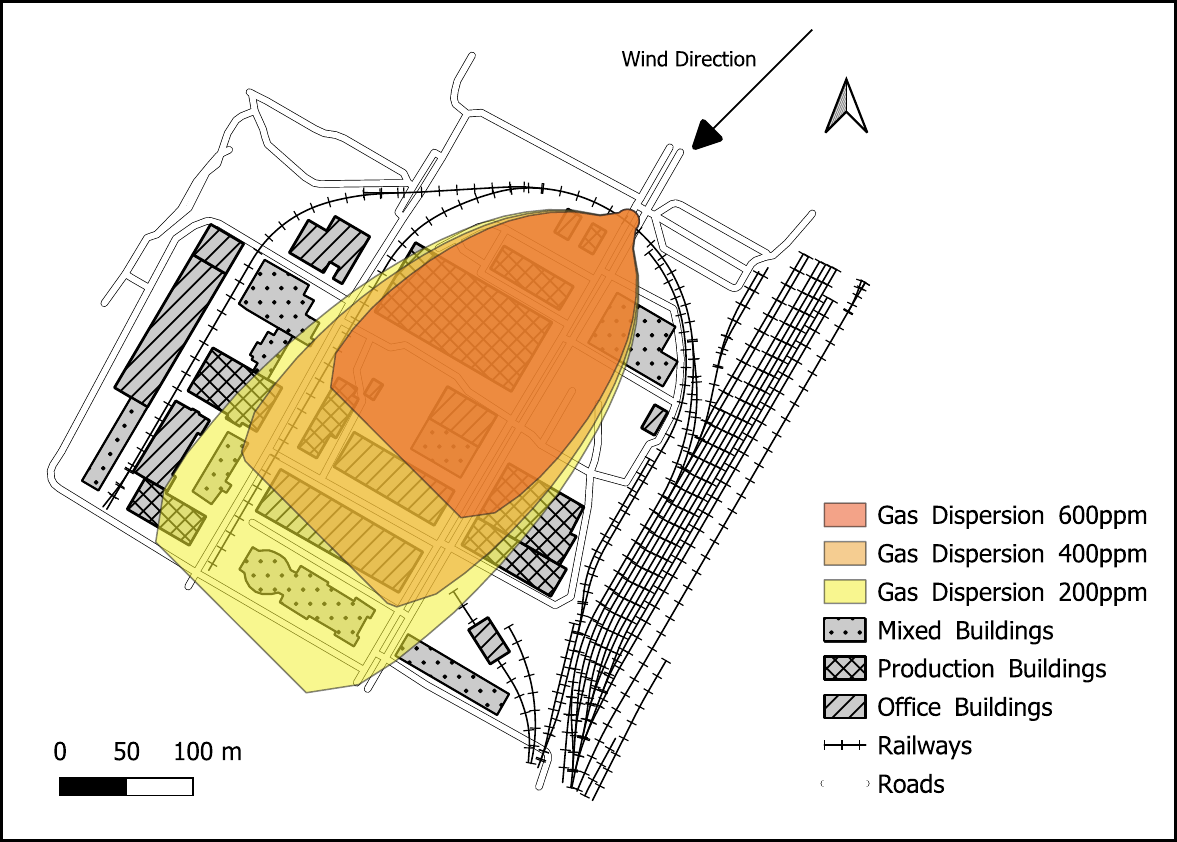}
	\caption{Simulation of the gas dispersion emerging from the critical junction including three threat levels characterised by different steady-state gas concentrations.}	\label{gas_dis}
\end{figure}

In the case study, the accident of the tank wagon with a truck caused an opening in the tank wagon with a diameter of 5 inches. The tank wagon is fully loaded with a volume of 100 $m^{3}$. Wind is coming from north east. Fig. \ref{gas_dis} shows the three threat zones (600ppm, 400ppm, and 200ppm) on the chemical plant site emerging from the crossing at the northern edge of the site. \\

In order to calculate a probability distribution for node \textit{Critical Gas Dose around Building}, the concentration of gas in the atmosphere is converted into a probability of a critical gas dose using a probability unit function (see Eq. \ref{eqconsersion} and \ref{eqconsersion2}). Probability unit (probit) functions are used to formulate a relationship between the criticality for a human being to the exposure of toxic substances \citep{Schubach1997, james2015simplified}. In addition to the concentration of gas, the time of exposure \textit{t} is required. The variables \textit{a}, \textit{b}, and \textit{n} are fitted constants, each specific for one type of toxic substance. For chlorine, $a=-8.29$, $b=0.92$, and $n=2$. The probit value $Y$ can be transferred into a probability of criticality using probit tables \citep{james2015simplified}. 

\begin{equation} \label{eqconsersion}
	Y=a+b\ln(Dose)
\end{equation} 

Dose is considered as integrated concentration of chemical exposure at a given point over a specific time \citep{james2015simplified}.
\begin{equation} \label{eqconsersion2}
	Dose=\int_0^{t}C^{n}dt
\end{equation} 

Applying the probit function for a chlorine exposure with e.g. $t=25min$ and $C_{1}=600$ppm, $C_{2}=400$ppm, and $C_{3}=200$ppm results in a probability of criticality of $90\%$ for $C_{1}$, $80\%$ for $C_{2}$, and $70\%$ for $C_{3}$. 

\section{Case Study Operation Phase}
\subsection{Single Building}
\begin{figure}[!h]
	\centering
	\includegraphics[width=10cm]{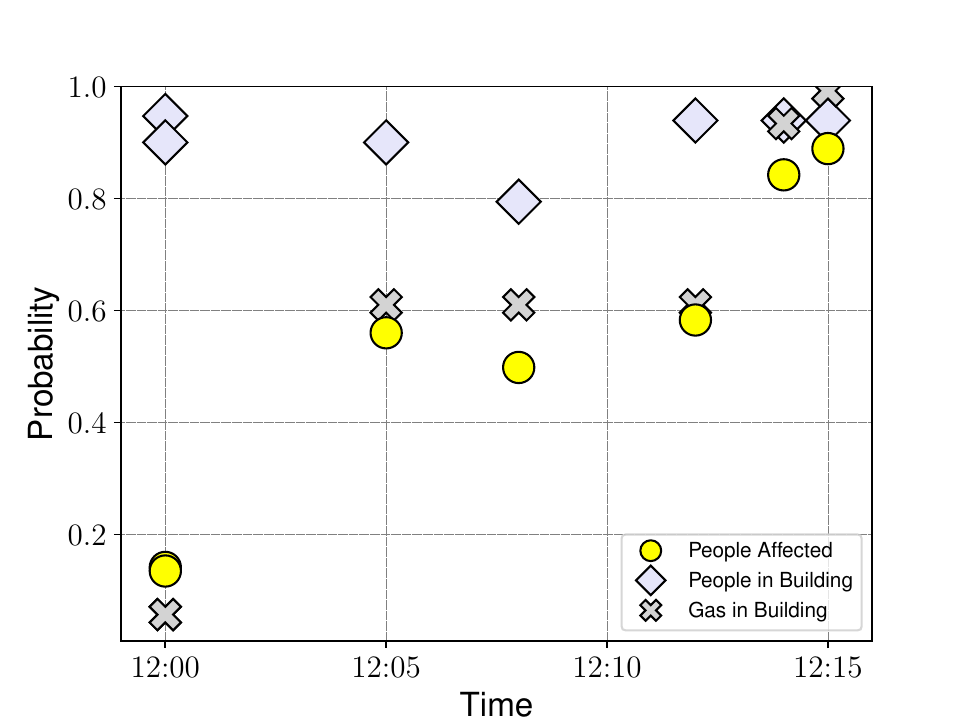}
	\caption{Probability of nodes \textit{People in Building Affected}, \textit{People in Building}, and \textit{Gas in Building} being in state \textit{True} for multiple time steps listed in Table \ref{sequence_one}.}	\label{bn_obs}
\end{figure}

First, the application of our \textit{ERIMap} method in the operation phase is illustrated for a single building on the chemical plant site (building 17 in Fig. \ref{case_study}). For this building, the associated BN is updated in accordance with an example sequence of observations (see Table \ref{sequence_one}). Based on the dynamically updated BN, the method helps to assess the probability that affected people are present in the building at each point in time (Fig. \ref{bn_obs}).\\

The time of day (12:00am) and building use (production building) are known instantly and are considered as hard evidence. Processing these observations results in a high probability for the presence of people (90\%) and a lower probability for people being affected (13\%). The probability of the dispersion of gas into the building is unaffected by this observation, i.e. it shows the prior probability (6\%). Next, the simulation of gas dispersion around the buildings is available (12:05am). Once the gas dispersion layer in the GIS shows an overlap with a building in the building layer, the observation obtained from the simulation is considered for that building and node \textit{Critical Gas Dosis around Building} (abbreviated as C.G.D. ar. Building). If a building shows an overlap with several gas dispersion layers, the one with the highest gas concentration is considered. Since building 17 shows an overlap with all three gas dispersion layers (see Fig. \ref{gas_dis}), a gas concentration of $600$ppm is assumed in the observation. In order to apply Eq. \ref{eqconsersion2}, an exposure time of 15 minutes is used under the assumption that the gas dispersion started at 11:50pm. Processing the observation obtained from this simulation, the probability of people in this building being affected jumps to 56\% while the probability of node \textit{Gas in Building} reaches 61\%. Next, an observation provided by a civilian of $RS_{1}$ is available (12:08am) stating that no people are in the building, followed by an observation by a second civilian (12:12am) of the same $RS_{1}$ stating the opposite. Due to the implementation of the regret function introduced in Section \ref{single-state}, the observation stating that people are present in the building is considered with a higher weight. After processing these contradictory observations, the probability of the presence of people is 94\%. At 12:14am, an observation by a gas sensor in the building is available, that provides evidence with an accuracy of 90\%. One minute later, a second gas sensor with the same accuracy provides an additional observation. Processing these two observations results in a probability of 99\% for node \textit{Gas in Building} and 89\% for node \textit{People in Building Affected}. 

\begin{table}[!h]
    \centering
    \caption{Example sequence of observations for building 17.}
    \label{sequence_one}
    \begin{tabular}{c|c|c|c|c}
    \toprule
        (1) Node & (2) State(s) & (3) Source (RS) & (4) Building(s) & (5) Time \\ \midrule
        \textit{Time of Day} & \textit{6am-6pm} & Clock ($RS_{3}$) & 17 & 12:00 am \\
        \textit{Building Type} & \textit{Production} & GIS Layer ($RS_{3}$) & 17 & 12:00 am \\
        \textit{C.G.D. ar. Building} & $P(0.8,0.2) $ & Simulation ($RS_{3}$) & 17 & 12:05 am \\
        \textit{People in Building} & \textit{False} & Civilian ($RS_{1}$) & 17 & 12:08 am \\
        \textit{People in Building} & \textit{True} & Civilian ($RS_{1}$) & 17 & 12:12 am \\
        \textit{C.G.D. in Building} & $L(0.9,0.1) $ & Gas Sensor ($RS_{3}$) & 17 & 12:14 am \\
        \textit{C.G.D. in Building} & $L(0.9,0.1) $ & Gas Sensor ($RS_{3}$) & 17 & 12:15 am \\\bottomrule
    \end{tabular}
\end{table}

\subsection{Results for all Buildings}
In order to compute the results for all buildings, two scenarios are designed that share the same gas dispersion at the chemical plant (see Fig. \ref{gas_dis}), but differ in the time of day and outcome of the situation. The two scenarios are constructed to highlight different aspects of the proposed method. The results show the probability of node \textit{People in Building Affected} being in state \textit{True} for each individual building at different time steps (four time steps in scenario 1 and six time steps in scenario 2). At each time step, new observations are available for multiple buildings. In the following, for each scenario, a brief summary is introduced, a table that includes the sequences of observations is provided, and results are outlined. \\

\subsubsection{Scenario 1}
\begin{figure}[!h]
	\centering
	\includegraphics[width=10cm]{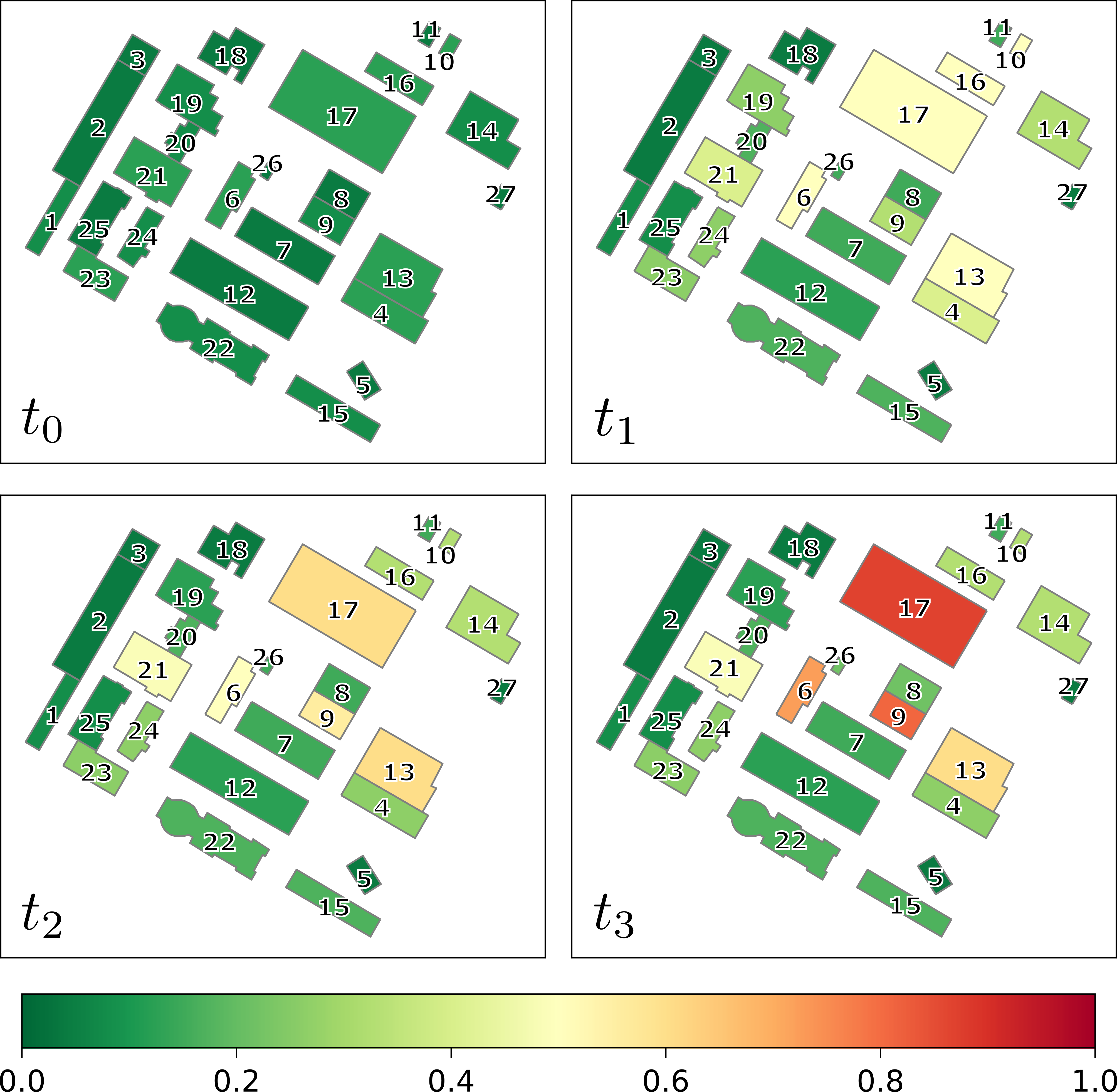}
	\caption{Probability of node \textit{People in Building Affected} being in state \textit{True} for each building at four time steps of scenario 1 (see Table\ref{seq1}).} 
    \label{results1}
\end{figure}

Scenario $1$ takes place during the night shift. At the initial time step $t_{0}$, the time of day and building types obtained from the GIS are available as observations. The results of $t_{0}$ show the resulting probability that differs between 4\% (for office building), 8\% (for mixed use buildings), and 12\% (for production buildings) (see Fig. \ref{results1}). At the next time step $t_{1}$, the observations by the gas dispersion simulation (see Fig. \ref{gas_dis}) become available. Here again, a gas exposure of 15 minutes is assumed as a first estimate resulting in the probability distributions for soft evidence for node \textit{Critical Gas Dose Around Building} shown in Table \ref{seq1}. The results show that given these observations, production buildings being located in the area of the simulated highest gas concentration (e.g. building 17) show the highest probability with 50\% (see Fig. \ref{results1} ). In the next time step ($t_{2}$), virtual evidence provided by individuals of $RS_{2}$ becomes available, stating that there are people in some buildings and that there are no people in another group of buildings. As displayed in Fig. \ref{results1} at time step $t_{2}$, buildings 9, 13, and 17 show the highest probability with over 50\%. At the last time step $t_{3}$, virtual evidence for node \textit{Critical Gas Dose in Building} provided by gas sensors with an accuracy of $90\%$ becomes available for multiple buildings. After processing these observations, building 17 shows the highest probability (84\%), followed by building 9 (81\%) and 6 (72\%).

\begin{table}[!h]
    \centering
    \caption{Sequence of observations of scenario 1.}
    \label{seq1}
    \begin{tabular}{c|c|c|c|c}
    \toprule
        (1) Node & (2) State(s) & (3) Source (RS) & (4) Buildings & (5) Time \\ \midrule
        \textit{Time of Day} & \textit{6pm-6am} & Clock ($RS_{3}$) & all & $t_{0}$ \\
        \textit{Building Type} & \textit{Office} & GIS Layer ($RS_{3}$) & 2, 3, 5, 7, 8, 11, 12, 18, 25, 26, 27 & $t_{0}$ \\
        \textit{Building Type} & \textit{Production} & GIS Layer ($RS_{3}$) & 4, 6, 10, 13, 16, 17, 21, 23 & $t_{0}$ \\
        \textit{Building Type} & \textit{Mixed} & GIS Layer ($RS_{3}$) & 1, 9, 14, 15, 19, 20, 22, 24 & $t_{0}$ \\
        
        \textit{C.G.D. ar. Building} & $P(0.3,0.7) $ & Simulation ($RS_{3}$) & 15, 20, 22, 23, 25 & $t_{1}$ \\
        \textit{C.G.D. ar. Building} & $P(0.6,0.4) $ & Simulation ($RS_{3}$) & 4, 12, 19, 21, 24 & $t_{1}$ \\
        \textit{C.G.D. ar. Building} & $P(0.8,0.2) $ & Simulation ($RS_{3}$) & 6, 7, 8, 9, 10, 11, 13, 14, 16, 17, 26 & $t_{1}$\\
        
        \textit{People in Building} & True & Human ($RS_{2}$) & 9,13,17,21 & $t_{2}$ \\
        \textit{People in Building} & False & Human ($RS_{2}$) & 4,10,16,19 & $t_{2}$ \\
        
        \textit{Gas in Building} & L(0.9,0.1) & Gas Sensor ($RS_{3}$) & 6,8,9,17,26 & $t_{3}$ \\\bottomrule
    \end{tabular}
\end{table}

\subsubsection{Scenario 2}
\begin{figure}[!h]
	\centering
	\includegraphics[width=15cm]{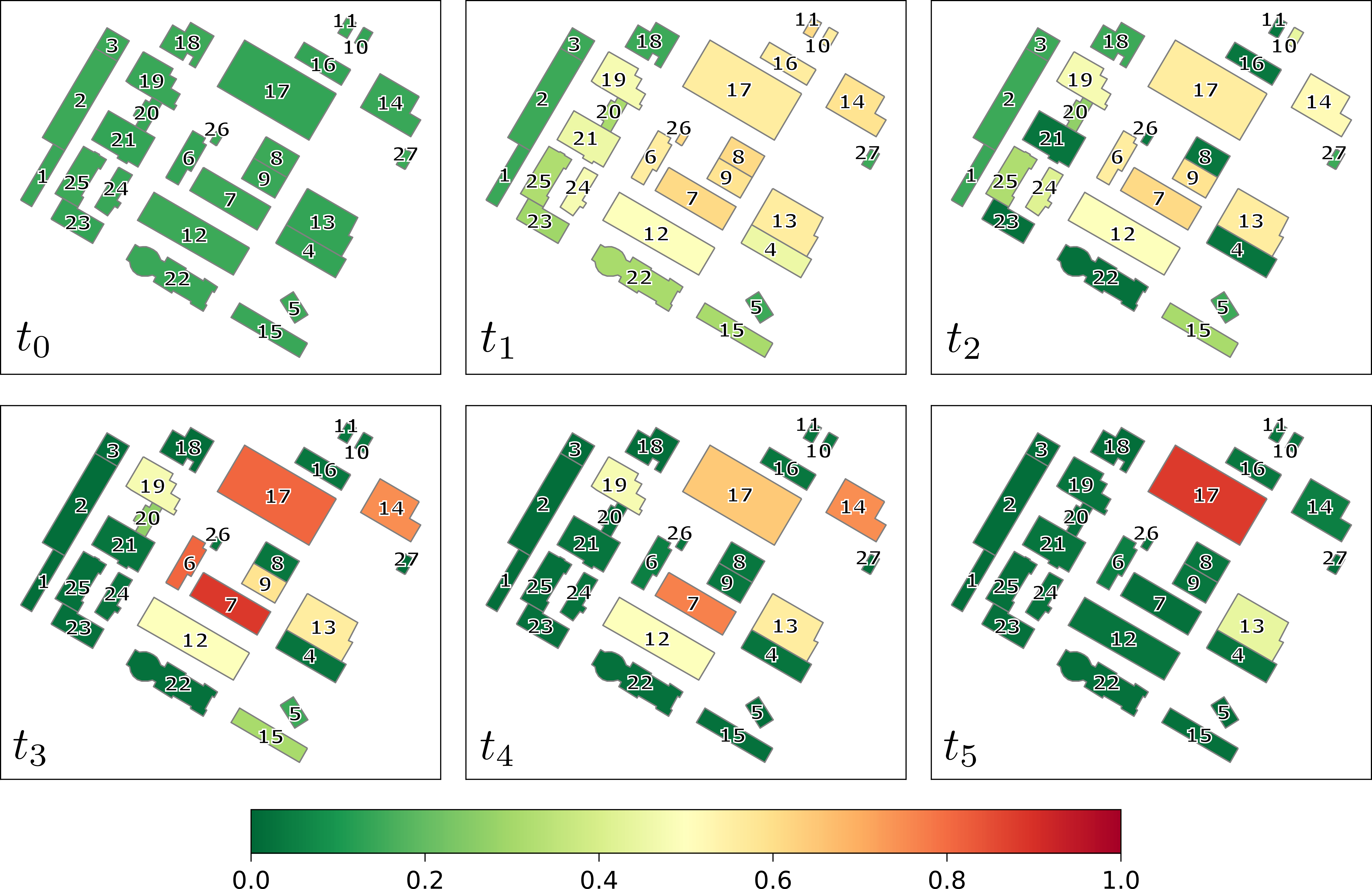}
	\caption{Probability of node \textit{People in Building Affected} being in state \textit{True} for each building at six time steps of scenario 2 (see Table \ref{seq2}).} 
    \label{results2}
\end{figure}

Scenario $2$ takes place during the day shift. At $t_{0}$, the time of day and building types obtained from the GIS are available as observations. The probability differs between 14\% (for office and mixed use buildings) and 15\% (for production buildings) (see Fig. \ref{results2}). At $t_{1}$, the observations provided by the gas dispersion simulation (see Fig. \ref{gas_dis}) becomes available (same as in scenario 1). Illustrated in Fig. \ref{results2} at $t_{1}$, the probability for the respective buildings varies between 14\% (e.g. building 1) and 61\% (e.g. building 7). At the next time step $t_{2}$, hard evidence provided by officials, i.e. humans of $RS_{3}$, becomes available stating that multiple buildings are evacuated, i.e. node \textit{People in Building} is in state \textit{False}. Additionally, observations by other individuals of $RS_{2}$ become available also stating that no people are in multiple other buildings. Buildings that are evacuated show a maximum probability of 3\% (e.g. building 8) and thus stand out clearly in Fig. \ref{results2} at $t_{2}$. At $t_{3}$, sensor information from gas sensors with an accuracy of 90\% becomes available, providing observations that include virtual evidence for multiple buildings. At the same time step, more buildings are evacuated and thus observations by officials of $RS_{3}$ become available. Buildings that are not yet evacuated can quickly be identified (see Fig. \ref{results2}). After this time step, building 7 shows the highest probability of node \textit{People in Building Affected} being in state \textit{True} with a probability of 89\%. At $t_{4}$, again, more buildings are officially evacuated and two observations become available each stating that no people are affected in building 7 and 17. These observations are provided by humans of $RS_{1}$ resulting in a decrease of probability for those two buildings. At the last time step $t_{5}$, more buildings are evacuated, an observation becomes available stating that affected people have been sighted in building 17, and an additional observation is provided by a human of $RS_{2}$ stating that no people are in building 13. At this time step, only building 13 and 17 are not yet evacuated with building 17 having a significantly higher probability of affected people compared to building 13.

\begin{table}[!h]
    \centering
    \caption{Sequence of observations of scenario 2.}
    \label{seq2}
    \begin{tabular}{c|c|c|c|c}
    \toprule
        (1) Node & (2) State(s) & (3) Source (RS) & (4) Buildings & (5) Time \\ \midrule
        \textit{Time of Day} & \textit{6pm-6am} & Clock ($RS_{3}$) & all & $t_{0}$ \\
        \textit{Building Type} & \textit{Office} & GIS Layer ($RS_{3}$) & 2, 3, 5, 7, 8, 11, 12, 18, 25, 26, 27 & $t_{0}$ \\
        \textit{Building Type} & \textit{Production} & GIS Layer ($RS_{3}$) & 4, 6, 10, 13, 16, 17, 21, 23 & $t_{0}$ \\
        \textit{Building Type} & \textit{Mixed} & GIS Layer ($RS_{3}$) & 1, 9, 14, 15, 19, 20, 22, 24 & $t_{0}$ \\
        
        \textit{C.G.D. ar. Building} & $P(0.3,0.7) $ & Simulation ($RS_{3}$) & 15, 20, 22, 23, 25 & $t_{1}$ \\
        \textit{C.G.D. ar. Building} & $P(0.6,0.4) $ & Simulation ($RS_{3}$) & 4, 12, 19, 21, 24 & $t_{1}$ \\
        \textit{C.G.D. ar. Building} & $P(0.8,0.2) $ & Simulation ($RS_{3}$) & 6, 7, 8, 9, 10, 11, 13, 14, 16, 17, 26 & $t_{1}$\\
        
        \textit{People in Building} & False & Human ($RS_{3}$) & 4,8,11,16,21,22,23,26 & $t_{2}$ \\               
        \textit{People in Building} & False & Human ($RS_{2}$) & 10,14,20,24 & $t_{2}$ \\

        \textit{Gas in Building} & L(0.9,0.1) &  Gas Sensor ($RS_{3}$) & 6,7,14,17 & $t_{3}$ \\        
        \textit{People in Building} & False & Human ($RS_{3}$) & 1,2,3,24,25,27,10,18 & $t_{3}$ \\

        \textit{People in Building} & False & Human ($RS_{3}$) & 5,6,9,15,20 & $t_{4}$ \\
        \textit{People Affected} & False & Human ($RS_{1}$) & 7,17 & $t_{4}$ \\

        \textit{People Affected} & True & Human ($RS_{2}$) & 17 & $t_{5}$ \\
        \textit{People in Building} & False & Human ($RS_{3}$) & 7,12,14,19 & $t_{5}$ \\
        \textit{People in Building} & False & Human ($RS_{2}$) & 13 & $t_{5}$
        \\\bottomrule
    \end{tabular}
\end{table}

\section{Discussion and Conclusion}
In this paper, we introduced \textit{ERIMap}, a novel Bayesian network-based method for supporting situation awareness tailored to the specific information-scape in emergency response. This specific information-scape can well be expressed via six requirements which have served as guiding principles for the design of the \textit{ERIMap} method. In accordance with these requirements, the \textit{ERIMap} method is capable of deriving insights about an ongoing situation by processing information which is incomplete (R1: process \textit{incomplete} information), which potentially stems from diverse sources (R2: process information from \textit{diverse sources}), which contains uncertainty (R3: process \textit{uncertain} information) and potentially contradictory observations (R4: process \textit{contradictory} information), which evolves dynamically in time (R5: process \textit{dynamic} information) and which is spatially distributed (R6: process \textit{spatial} information).  \\

Regarding the first requirement (R1: process \textit{incomplete} information), using a BN as the core of the method allows to incorporate pre-existing knowledge about important system variables and their interrelations into the assessment of an actual emergency situation (e.g. see \cite{Waal2007}). In this way, the method can draw inferences about an ongoing emergency situation based on incomplete information. For instance, already in the initial stages of the case study, the \textit{ERIMap} method provides an estimation on where the presence of affected people can be expected, based solely on information regarding the building types, the time of day, and the estimated gas dispersion. \\

Regarding the second requirement (R2: process information from \textit{diverse sources}), we have demonstrated how observations provided by different individuals, different sensors, GIS layers and simulations inform the \textit{ERIMap} method. Considering multiple sources increases the amount of information which can be utilised in the assessment of the situation, but without creating cognitive overload that is typically hampering emergency management \citep{van2016improving}. The \textit{ERIMap} method condenses and combines all single pieces of information into a single output; in the case study, the probability of affected people being in a building. \\

In emergencies many observations contain a certain degree of uncertainty (R3: process \textit{uncertain} information). However, in the literature, there are only few papers that apply uncertain evidence in BNs (e.g., \cite{Giordano2013}). According to \cite{Munk2023}, this might be due to a lack of consensus on which type of uncertain evidence should be applied in which case. A core feature of our \textit{ERIMap} method is a classification scheme which selects the 'right' type of evidence based on the reliability of the observation source (quantified by reliability scores) and the precision of the reported observation (a specific indication or a likelihood). Every piece of information can then be fed into the BN, taking into account the degree and type of uncertainty associated with it. In short, the \textit{ERIMap} method provides an evidence-specific protocol for hard, for soft and for virtual evidence. \\ 

The utilisation of uncertain evidence is also important for addressing the fourth requirement (R4: process \textit{conflicting} information). Our \textit{ERIMap} method contains two tools for dealing with conflicting observations, both of which rely on the utilisation of uncertain evidence: (1) reliability scores which favour observation sources which are considered more trustworthy and (2) a regret function which favours observations which report critical states ("there are people in building A") over observations which report non-critical states ("there are no people in building A"). While reliability scores and regret functions cannot dissolve the issue of conflicting observations, they can still provide a structured way for dealing with this type of noisy input. In particular, they allow to shift considerations on how to deal with conflicting information from the time-critical operation phase (during the emergency) to the non-time-critical preparation phase (prior to the emergency). \\ 

In the case study, we have demonstrated that the \textit{ERIMap} method is capable of providing a dynamically evolving (R5: process \textit{dynamic} information) and spatially resolved (R6: process \textit{spatial} information) picture of the current state of belief about an emergency situation. Due to the established protocol for translating observations into evidence, every new observation can directly be utilised to update the BN-based assessment of the current situation. Furthermore, the use of area-specific BNs in our \textit{ERIMap} method allows to cover the spatial dimension of emergencies. The implementation of the area specification is kept quite simple. It is realised by initially assigning duplicates of the same BN to every subarea within the study site. The area-specific output is then obtained by feeding each BN with area-specific evidence. While we assume that this simple approach is sufficient for many applications, it should be noted that it can easily be adapted to more complex demands, for instance, one could use different BNs for indoor and outdoor areas or for different types of buildings. In the same manner, the spatial resolution of the \textit{ERIMap} can be adjusted according to the demands of the addressed emergency response team. \\

Besides fulfilling all requirements, a final advantage of BNs is that, due to their graphical structure, they represent explainable models whose dynamics are comprehensible, even for people who are not familiar with their technical details \citep{Schaberreiter2014}. Using an explainable model facilitates engaging experts and potential users in the development and validation of the model. In this way, the model can be established even if the data situation is not satisfactory. In addition, the graphical nature of the model facilitates adjustments to the preferences of emergency responders. What is more, keeping the process of drawing inferences from observations transparent enhances the acceptance by potential users - an aspect that is crucial in emergency response \citep{Weidinger2020}. Especially in domains where decisions have a direct impact on people - such as in emergency response - there is a certain reluctance to trust black box models \citep{Waal2022} and ease of interpretation is vital \citep{lee2022roadmap}. \\

A main limitation of the \textit{ERIMap} method is that, for setting up the BN at its core, the method relies on structural knowledge about the emergency in question which exists prior to the actual emergency situation which it should help to assess. Importantly, the value which the method adds to this assessment ultimately depends on the quality of this pre-existing knowledge. This implies that eliciting this knowledge during the preparation phase is of utmost importance for the successful application of the \textit{ERIMap} method. While we have highlighted the importance of engaging users in this process, we have not covered how their knowledge can best be elicited for the use in the \textit{ERIMap} method. While this is an aspect that has already been addressed for BNs (e.g. \cite{Hassall2019} or \cite{Morris2014}), it will nevertheless be important to establish a corresponding procedure which is specific to the \textit{ERIMap} method. Another implication is that the method is particularly well suited for rather common or expectable types of emergencies in well-known (and structured) territories. The \textit{ERIMap} method thus seems promising for supporting the operation of emergency response teams in emergencies which are considered particularly relevant in specific facilities, for instance, for gas leakages on chemical plant sites (our case study), for fires in office buildings or airports, or for floods in cities situated near rivers or coasts. To what extent the \textit{ERIMap} method can also contribute to assessing important variables of an unexpected or unforeseen emergency situation is less clear and should be further explored in future studies. \\

A technical aspect which has not yet been covered in the \textit{ERIMap} method are mechanisms for performing belief updates in the BNs which are not directly tied to newly available observations but which rely on presumably predictable inner system dynamics. For instance, in future work, it could be insightful to consider the movement of people (e.g., evacuation of buildings), the dispersion of hazardous material (like gas or smoke) or the increasing criticality of being exposed to such material using dynamic Bayesian networks \citep{Murphy2002} and spatially interlinked area-specific BNs. However, in how far belief updates based on presumed dynamics can and should be utilised to assess real emergency situations needs to be evaluated in dialogue with potential users.\\

One limitation of this paper is that we still need to empirically test the impact of the \textit{ERIMap} on situation awareness. First feedback from potential users (e.g., members of the Henkel fire brigade) is thoroughly positive. However, in order to verify its positive impact on situation awareness, future work should include empirical testing and validation of the method in practice. To this end, the method should be established in a real setting, be integrated in the existing emergency response protocols and then be evaluated in training exercises with responsible practitioners. For the further development of the \textit{ERIMap} method this implies that future work should focus on facilitating the transfer of the \textit{ERIMap} method into practice. In particular, future research should focus on (1) how expert knowledge can best be compiled in the preparation of the \textit{ERIMap} method; on (2) developing a user interface which facilitates a quick and straightforward injection of observations (including the five pieces of information, e.g., the reliability score) and which displays the results in an interactive map; and on (3) empirical and experimental testing and measuring the impact of the \textit{ERIMap} on situation awareness, e.g., in training sessions or serious games. 


\begin{thebibliography}{}

\bibitem[\protect\citeauthoryear{Abdalla}{Abdalla}{2016}]{Abdalla2016}
Abdalla, R. (2016, December).
\newblock Evaluation of spatial analysis application for urban emergency management.
\newblock {\em SpringerPlus\/}~{\em 5\/}(1), 2081--2091.

\bibitem[\protect\citeauthoryear{Ankan and Panda}{Ankan and Panda}{2015}]{ankan2015pgmpy}
Ankan, A. and A.~Panda (2015).
\newblock pgmpy: Probabilistic graphical models using python.
\newblock In {\em Proceedings of the 14th Python in Science Conference (SCIPY 2015)}. Citeseer.

\bibitem[\protect\citeauthoryear{Avvenuti, Cresci, Del~Vigna, Fagni, and Tesconi}{Avvenuti et~al.}{2018}]{Avvenuti2018}
Avvenuti, M., S.~Cresci, F.~Del~Vigna, T.~Fagni, and M.~Tesconi (2018, October).
\newblock {CrisMap}: a {Big} {Data} {Crisis} {Mapping} {System} {Based} on {Damage} {Detection} and {Geoparsing}.
\newblock {\em Information Systems Frontiers\/}~{\em 20\/}(5), 993--1011.

\bibitem[\protect\citeauthoryear{Brusca, Famoso, Lanzafame, Mauro, Garrano, and Monforte}{Brusca et~al.}{2016}]{Brusca2016}
Brusca, S., F.~Famoso, R.~Lanzafame, S.~Mauro, A.~M.~C. Garrano, and P.~Monforte (2016, November).
\newblock Theoretical and {Experimental} {Study} of {Gaussian} {Plume} {Model} in {Small} {Scale} {System}.
\newblock {\em Energy Procedia\/}~{\em 101}, 58--65.

\bibitem[\protect\citeauthoryear{Chan and Darwiche}{Chan and Darwiche}{2005}]{Chan2005}
Chan, H. and A.~Darwiche (2005, March).
\newblock On the revision of probabilistic beliefs using uncertain evidence.
\newblock {\em Artificial Intelligence\/}~{\em 163\/}(1), 67--90.

\bibitem[\protect\citeauthoryear{Comes, Van~de Walle, and Van~Wassenhove}{Comes et~al.}{2020}]{comes2020}
Comes, T., B.~Van~de Walle, and L.~Van~Wassenhove (2020, June).
\newblock The coordination-information bubble in humanitarian response: theoretical foundations and empirical investigations.
\newblock {\em Production and Operations Management\/}~{\em 29\/}(11), 2484--2507.

\bibitem[\protect\citeauthoryear{Comes, Wijngaards, Maule, Allen, and Schultmann}{Comes et~al.}{2012}]{Comes2012}
Comes, T., N.~Wijngaards, J.~Maule, D.~Allen, and F.~Schultmann (2012, March).
\newblock Scenario reliability assessment to support decision makers in situations of severe uncertainty.
\newblock In {\em 2012 {IEEE} {International} {Multi}-{Disciplinary} {Conference} on {Cognitive} {Methods} in {Situation} {Awareness} and {Decision} {Support}}, pp.\  30--37.
\newblock ISSN: 2379-1675.

\bibitem[\protect\citeauthoryear{Comes, Wijngaards, and Schultmann}{Comes et~al.}{2012}]{Comes.2012}
Comes, T., N.~Wijngaards, and F.~Schultmann (2012, April).
\newblock Efficient scenarios updating in emergency management.
\newblock In {\em Proceedings of the 9th International Conference on Information Systems for Crisis Response and Management (ISCRAM), April 22-25, 2012, Vancouver, Canada}, pp.\  167.

\bibitem[\protect\citeauthoryear{Comes, Wijngaards, and Van~de Walle}{Comes et~al.}{2015}]{Comes2015}
Comes, T., N.~Wijngaards, and B.~Van~de Walle (2015, August).
\newblock Exploring the future: {Runtime} scenario selection for complex and time-bound decisions.
\newblock {\em Technological Forecasting and Social Change\/}~{\em 97}, 29--46.

\bibitem[\protect\citeauthoryear{de~Waal and Joubert}{de~Waal and Joubert}{2022}]{Waal2022}
de~Waal, A. and J.~W. Joubert (2022, December).
\newblock Explainable {Bayesian} networks applied to transport vulnerability.
\newblock {\em Expert Systems with Applications\/}~{\em 209}, 118348.

\bibitem[\protect\citeauthoryear{Dlamini}{Dlamini}{2011}]{Dlamini2011}
Dlamini, W.~M. (2011, June).
\newblock Application of {Bayesian} networks for fire risk mapping using {GIS} and remote sensing data.
\newblock {\em GeoJournal\/}~{\em 76\/}(3), 283--296.

\bibitem[\protect\citeauthoryear{Dlamini, Simelane, and Nhlabatsi}{Dlamini et~al.}{2022}]{Dlamini2022}
Dlamini, W. M.~D., S.~P. Simelane, and N.~M. Nhlabatsi (2022, February).
\newblock Bayesian network-based spatial predictive modelling reveals {COVID}-19 transmission dynamics in {Eswatini}.
\newblock {\em Spatial Information Research\/}~{\em 30\/}(1), 183--194.

\bibitem[\protect\citeauthoryear{Druzdzel and van~der Gaag}{Druzdzel and van~der Gaag}{2000}]{Druzdzel2000}
Druzdzel, M. and L.~van~der Gaag (2000, July).
\newblock Building probabilistic networks: "{Where} do the numbers come from?" guest editors' introduction.
\newblock {\em IEEE Transactions on Knowledge and Data Engineering\/}~{\em 12\/}(4), 481--486.

\bibitem[\protect\citeauthoryear{Endsley}{Endsley}{1995}]{Endsley1995}
Endsley, M.~R. (1995, March).
\newblock Toward a theory of situation awareness in dynamic systems.
\newblock {\em Human factors\/}~{\em 37\/}(1), 32--64.

\bibitem[\protect\citeauthoryear{Fathi, Thom, Koch, Ertl, and Fiedrich}{Fathi et~al.}{2020}]{Fathi2020}
Fathi, R., D.~Thom, S.~Koch, T.~Ertl, and F.~Fiedrich (2020, July).
\newblock {VOST}: {A} case study in voluntary digital participation for collaborative emergency management.
\newblock {\em Information Processing \& Management\/}~{\em 57\/}(4), 102174.

\bibitem[\protect\citeauthoryear{Geiß, Priesmeier, Aravena~Pelizari, Soto~Calderon, Schoepfer, Riedlinger, Villar~Vega, Santa~María, Gómez~Zapata, Pittore, So, Fekete, and Taubenböck}{Geiß et~al.}{2022}]{Geiss2022}
Geiß, C., P.~Priesmeier, P.~Aravena~Pelizari, A.~R. Soto~Calderon, E.~Schoepfer, T.~Riedlinger, M.~Villar~Vega, H.~Santa~María, J.~C. Gómez~Zapata, M.~Pittore, E.~So, A.~Fekete, and H.~Taubenböck (2022, November).
\newblock Benefits of global earth observation missions for disaggregation of exposure data and earthquake loss modeling: evidence from {Santiago} de {Chile}.
\newblock {\em Natural Hazards\/}~{\em 119\/}(1), 779--804.

\bibitem[\protect\citeauthoryear{Giordano, D'Agostino, Apollonio, Lamaddalena, and Vurro}{Giordano et~al.}{2013}]{Giordano2013}
Giordano, R., D.~D'Agostino, C.~Apollonio, N.~Lamaddalena, and M.~Vurro (2013, December).
\newblock Bayesian {Belief} {Network} to support conflict analysis for groundwater protection: {The} case of the {Apulia} region.
\newblock {\em Journal of Environmental Management\/}~{\em 115}, 136--146.

\bibitem[\protect\citeauthoryear{Guo, Guan, and Yan}{Guo et~al.}{2023}]{Guo2023}
Guo, K., M.~Guan, and H.~Yan (2023, July).
\newblock Utilising social media data to evaluate urban flood impact in data scarce cities.
\newblock {\em International Journal of Disaster Risk Reduction\/}~{\em 93}, 103780.

\bibitem[\protect\citeauthoryear{Hao, Xu, Zhao, and Fujita}{Hao et~al.}{2018}]{Hao2018}
Hao, Z., Z.~Xu, H.~Zhao, and H.~Fujita (2018, August).
\newblock A {Dynamic} {Weight} {Determination} {Approach} {Based} on the {Intuitionistic} {Fuzzy} {Bayesian} {Network} and {Its} {Application} to {Emergency} {Decision} {Making}.
\newblock {\em IEEE Transactions on Fuzzy Systems\/}~{\em 26\/}(4), 1893--1907.

\bibitem[\protect\citeauthoryear{Hassall, Dailey, Zawadzka, Milne, Harris, Corstanje, and Whitmore}{Hassall et~al.}{2019}]{Hassall2019}
Hassall, K.~L., G.~Dailey, J.~Zawadzka, A.~E. Milne, J.~A. Harris, R.~Corstanje, and A.~P. Whitmore (2019, December).
\newblock Facilitating the elicitation of beliefs for use in {Bayesian} {Belief} modelling.
\newblock {\em Environmental Modelling \& Software\/}~{\em 122}, 104539.

\bibitem[\protect\citeauthoryear{Huang, Liu, Wang, Yuan, and Chen}{Huang et~al.}{2022}]{Huang2022}
Huang, L., G.~Liu, Y.~Wang, H.~Yuan, and T.~Chen (2022, April).
\newblock Fire detection in video surveillances using convolutional neural networks and wavelet transform.
\newblock {\em Engineering Applications of Artificial Intelligence\/}~{\em 110}, 104737.

\bibitem[\protect\citeauthoryear{James}{James}{2014}]{james2015simplified}
James, M. (2014, June).
\newblock Simplified methods of using probit analysis in consequence analysis.
\newblock {\em Process Safety Progress\/}~{\em 34\/}(1), 58--63.

\bibitem[\protect\citeauthoryear{Javed, Norris, and Johnston}{Javed et~al.}{2011}]{Javed2011}
Javed, Y., T.~Norris, and D.~Johnston (2011, May).
\newblock Ontology-based inference to enhance team situation awareness in emergency management.
\newblock In L.~S. M.A.~Santos (Ed.), {\em 8th {International} {Conference} on {Information} {Systems} for {Crisis} {Response} and {Management}: {From} {Early}-{Warning} {Systems} to {Preparedness} and {Training}, {ISCRAM} 2011}, Lisbon. Information Systems for Crisis Response and Management, ISCRAM.

\bibitem[\protect\citeauthoryear{Johnson, Low-Choy, and Mengersen}{Johnson et~al.}{2012}]{Johnson2012}
Johnson, S., S.~Low-Choy, and K.~Mengersen (2012, August).
\newblock Integrating {Bayesian} networks and geographic information systems: {Good} practice examples.
\newblock {\em Integrated Environmental Assessment and Management\/}~{\em 8\/}(3), 473--479.

\bibitem[\protect\citeauthoryear{Jordahl, den Bossche, Fleischmann, Wasserman, McBride, Gerard, Tratner, Perry, Badaracco, Farmer, Hjelle, Snow, Cochran, Gillies, Culbertson, Bartos, Eubank, maxalbert, Bilogur, Rey, Ren, Arribas-Bel, Wasser, Wolf, Journois, Wilson, Greenhall, Holdgraf, Filipe, and Leblanc}{Jordahl et~al.}{2020}]{kelsey_jordahl_2020_3946761}
Jordahl, K., J.~V. den Bossche, M.~Fleischmann, J.~Wasserman, J.~McBride, J.~Gerard, J.~Tratner, M.~Perry, A.~G. Badaracco, C.~Farmer, G.~A. Hjelle, A.~D. Snow, M.~Cochran, S.~Gillies, L.~Culbertson, M.~Bartos, N.~Eubank, maxalbert, A.~Bilogur, S.~Rey, C.~Ren, D.~Arribas-Bel, L.~Wasser, L.~J. Wolf, M.~Journois, J.~Wilson, A.~Greenhall, C.~Holdgraf, Filipe, and F.~Leblanc (2020, July).
\newblock geopandas/geopandas: v0.8.1.

\bibitem[\protect\citeauthoryear{Lee, Comes, Finn, and Mostafavi}{Lee et~al.}{2022}]{lee2022roadmap}
Lee, C.-C., T.~Comes, M.~Finn, and A.~Mostafavi (2022).
\newblock Roadmap towards responsible ai in crisis resilience management.

\bibitem[\protect\citeauthoryear{Li, Lu, Ji, Fan, and Li}{Li et~al.}{2023}]{Li2023}
Li, B., J.~Lu, Y.~Ji, H.~Fan, and J.~Li (2023, August).
\newblock A dynamic emergency response decision-making method considering the scenario evolution of maritime emergencies.
\newblock {\em Computers \& Industrial Engineering\/}~{\em 182}, 109438.

\bibitem[\protect\citeauthoryear{Milana}{Milana}{2022}]{Milana2022}
Milana, E. (2022, November).
\newblock Soft robotics for infrastructure protection.
\newblock {\em Frontiers in Robotics and AI\/}~{\em 9}.

\bibitem[\protect\citeauthoryear{Morris, Oakley, and Crowe}{Morris et~al.}{2014}]{Morris2014}
Morris, D.~E., J.~E. Oakley, and J.~A. Crowe (2014, February).
\newblock A web-based tool for eliciting probability distributions from experts.
\newblock {\em Environmental Modelling and Software\/}~{\em 52}, 1--4.

\bibitem[\protect\citeauthoryear{Mrad, Delcroix, Maalej, Piechowiak, and Abid}{Mrad et~al.}{2012}]{Mrad2012}
Mrad, A.~B., V.~Delcroix, M.~A. Maalej, S.~Piechowiak, and M.~Abid (2012).
\newblock Uncertain evidence in bayesian networks: Presentation and comparison on a simple example.
\newblock In {\em Communications in Computer and Information Science}, pp.\  39--48. Springer Berlin Heidelberg.

\bibitem[\protect\citeauthoryear{Mrad, Delcroix, Piechowiak, Leicester, and Abid}{Mrad et~al.}{2015}]{Mrad2015}
Mrad, A.~B., V.~Delcroix, S.~Piechowiak, P.~Leicester, and M.~Abid (2015, June).
\newblock An explication of uncertain evidence in bayesian networks: likelihood evidence and probabilistic evidence.
\newblock {\em Applied Intelligence\/}~{\em 43\/}(4), 802--824.

\bibitem[\protect\citeauthoryear{Mrad, Delcroix, Piechowiak, Maalej, and Abid}{Mrad et~al.}{2013}]{BenMrad2013}
Mrad, A.~B., V.~Delcroix, S.~Piechowiak, M.~A. Maalej, and M.~Abid (2013, April).
\newblock Understanding soft evidence as probabilistic evidence: {Illustration} with several use cases.
\newblock In {\em 2013 5th {International} {Conference} on {Modeling}, {Simulation} and {Applied} {Optimization} ({ICMSAO})}, pp.\  1--6.

\bibitem[\protect\citeauthoryear{Munk, Mead, and Wood}{Munk et~al.}{2023}]{Munk2023}
Munk, A., A.~Mead, and F.~Wood (2023, July).
\newblock Uncertain {Evidence} in {Probabilistic} {Models} and {Stochastic} {Simulators}.
\newblock pp.\  25486--25500. PMLR.

\bibitem[\protect\citeauthoryear{Murphy}{Murphy}{2002}]{Murphy2002}
Murphy, K. (2002, January).
\newblock {\em Dynamic Bayesian Networks: Representation, Inference and Learning}.
\newblock Ph.\ D. thesis.

\bibitem[\protect\citeauthoryear{Pan, Peng, and Ding}{Pan et~al.}{2006}]{Pan2006}
Pan, R., Y.~Peng, and Z.~Ding (2006, November).
\newblock Belief update in bayesian networks using uncertain evidence.
\newblock In {\em 2006 18th {IEEE} International Conference on Tools with Artificial Intelligence ({ICTAI}{\textquotesingle}06)}. {IEEE}.

\bibitem[\protect\citeauthoryear{Pearl}{Pearl}{1985}]{pearl1985bayesian}
Pearl, J. (1985, June).
\newblock Bayesian networks: A model of self-activated memory for evidential reasoning.
\newblock In {\em Proceedings of the 7th conference of the Cognitive Science Society, University of California, Irvine, CA, USA}, pp.\  15--17.

\bibitem[\protect\citeauthoryear{Pearl}{Pearl}{1988}]{Pearl1988}
Pearl, J. (1988, September).
\newblock {\em Probabilistic {Reasoning} in {Intelligent} {Systems}: {Networks} of {Plausible} {Inference}}.
\newblock Morgan Kaufmann.

\bibitem[\protect\citeauthoryear{Peng, Zhang, and Pan}{Peng et~al.}{2010}]{YUN2010}
Peng, Y., S.~Zhang, and R.~Pan (2010, October).
\newblock {BAYESIAN} {NETWORK} {REASONING} {WITH} {UNCERTAIN} {EVIDENCES}.
\newblock {\em International Journal of Uncertainty, Fuzziness and Knowledge-Based Systems\/}~{\em 18\/}(05), 539--564.

\bibitem[\protect\citeauthoryear{Radianti, Granmo, Sarshar, Goodwin, Dugdale, and Gonzalez}{Radianti et~al.}{2015}]{Radianti2015a}
Radianti, J., O.-C. Granmo, P.~Sarshar, M.~Goodwin, J.~Dugdale, and J.~J. Gonzalez (2015, September).
\newblock A spatio-temporal probabilistic model of hazard- and crowd dynamics for evacuation planning in disasters.
\newblock {\em Applied Intelligence\/}~{\em 42\/}(1), 3--23.

\bibitem[\protect\citeauthoryear{Ram\'irez-Agudelo, K\"opke, Guillouet, Schäfer-Frey, Engler, Mielniczek, and Sill~Torres}{Ram\'irez-Agudelo et~al.}{2021}]{RamirezAgudelo2021}
Ram\'irez-Agudelo, O.~H., C.~K\"opke, Y.~Guillouet, J.~Schäfer-Frey, E.~Engler, J.~Mielniczek, and F.~Sill~Torres (2021, September).
\newblock An {Expert}-{Driven} {Probabilistic} {Assessment} of the {Safety} and {Security} of {Offshore} {Wind} {Farms}.
\newblock {\em Energies\/}~{\em 14\/}(17), 102--108.

\bibitem[\protect\citeauthoryear{Ricci, Yang, Reniers, and Cozzani}{Ricci et~al.}{2024}]{Ricci2024}
Ricci, F., M.~Yang, G.~Reniers, and V.~Cozzani (2024, March).
\newblock Emergency response in cascading scenarios triggered by natural events.
\newblock {\em Reliability Engineering \& System Safety\/}~{\em 243}, 109820.

\bibitem[\protect\citeauthoryear{Rohr, Priesmeier, Tzavella, and Fekete}{Rohr et~al.}{2020}]{Rohr2020}
Rohr, A., P.~Priesmeier, K.~Tzavella, and A.~Fekete (2020, November).
\newblock System {Criticality} of {Road} {Network} {Areas} for {Emergency} {Management} {Services} - {Spatial} {Assessment} {Using} a {Tessellation} {Approach}.
\newblock {\em Infrastructures\/}~{\em 5}.

\bibitem[\protect\citeauthoryear{Schaberreiter, Bouvry, Röning, and Khadraoui}{Schaberreiter et~al.}{2014}]{Schaberreiter2014}
Schaberreiter, T., P.~Bouvry, J.~Röning, and D.~Khadraoui (2014).
\newblock Support {Tool} for a {Bayesian} {Network} {Based} {Critical} {Infrastructure} {Risk} {Model}.
\newblock In O.~Schuetze, C.~A. Coello~Coello, A.-A. Tantar, E.~Tantar, P.~Bouvry, P.~D. Moral, and P.~Legrand (Eds.), {\em {EVOLVE} - {A} {Bridge} between {Probability}, {Set} {Oriented} {Numerics}, and {Evolutionary} {Computation} {III}}, Studies in {Computational} {Intelligence}, Heidelberg, pp.\  53--75. Springer International Publishing.

\bibitem[\protect\citeauthoryear{Schubach}{Schubach}{1997}]{Schubach1997}
Schubach, S. (1997, September).
\newblock A measure of human sensitivity in acute inhalation toxicity.
\newblock {\em Journal of Loss Prevention in the Process Industries\/}~{\em 10\/}(5), 309--315.

\bibitem[\protect\citeauthoryear{Seixas, Zadrozny, Laks, Conci, and Muchaluat~Saade}{Seixas et~al.}{2014}]{Seixas2014}
Seixas, F.~L., B.~Zadrozny, J.~Laks, A.~Conci, and D.~C. Muchaluat~Saade (2014, August).
\newblock A {Bayesian} network decision model for supporting the diagnosis of dementia, alzheimer's disease and mild cognitive impairment.
\newblock {\em Computers in Biology and Medicine\/}~{\em 51}, 140--158.

\bibitem[\protect\citeauthoryear{Turoff, Chumer, Walle, and Yao}{Turoff et~al.}{2004}]{Turoff2004}
Turoff, M., M.~Chumer, B.~d. Walle, and X.~Yao (2004, January).
\newblock The {Design} of a {Dynamic} {Emergency} {Response} {Management} {Information} {System} ({DERMIS}).
\newblock {\em Journal of Information Technology Theory and Application (JITTA)\/}~{\em 5\/}(4).

\bibitem[\protect\citeauthoryear{Tzavella, Fekete, and Fiedrich}{Tzavella et~al.}{2018}]{Tzavella2018}
Tzavella, K., A.~Fekete, and F.~Fiedrich (2018, April).
\newblock Opportunities provided by geographic information systems and volunteered geographic information for a timely emergency response during flood events in {Cologne}, {Germany}.
\newblock {\em Natural Hazards\/}~{\em 91\/}(1), 29--57.

\bibitem[\protect\citeauthoryear{Valtorta, Kim, and Vomlel}{Valtorta et~al.}{2002}]{Valtorta2002}
Valtorta, M., Y.-G. Kim, and J.~Vomlel (2002, January).
\newblock Soft evidential update for probabilistic multiagent systems.
\newblock {\em International Journal of Approximate Reasoning\/}~{\em 29\/}(1), 71--106.

\bibitem[\protect\citeauthoryear{Van~de Walle, Brugghemans, and Comes}{Van~de Walle et~al.}{2016}]{van2016improving}
Van~de Walle, B., B.~Brugghemans, and T.~Comes (2016, November).
\newblock Improving situation awareness in crisis response teams: {An} experimental analysis of enriched information and centralized coordination.
\newblock {\em International Journal of Human-Computer Studies\/}~{\em 95}, 66--79.

\bibitem[\protect\citeauthoryear{van~de Walle and Turoff}{van~de Walle and Turoff}{2008}]{Walle2008}
van~de Walle, B. and M.~Turoff (2008).
\newblock Decision {Support} for {Emergency} {Situations}.
\newblock International {Handbooks} {Information} {System}, pp.\  39--63. Berlin, Heidelberg: Springer.

\bibitem[\protect\citeauthoryear{Vomlel}{Vomlel}{2004}]{Vomlel2004}
Vomlel, J. (2004, January).
\newblock Probabilistic reasoning with uncertain evidence.
\newblock {\em International Journal on Neural and Mass-Parallel Computing and Information Systems\/}~{\em 14\/}(5), 453--456.

\bibitem[\protect\citeauthoryear{Waal and Ritchey}{Waal and Ritchey}{2007}]{Waal2007}
Waal, A.~D. and T.~Ritchey (2007, December).
\newblock Combining morphological analysis and bayesian networks for strategic decision support.
\newblock {\em ORiON\/}~{\em 23\/}(2).

\bibitem[\protect\citeauthoryear{Weber, Medina-Oliva, Simon, and Iung}{Weber et~al.}{2012}]{Weber2012}
Weber, P., G.~Medina-Oliva, C.~Simon, and B.~Iung (2012, June).
\newblock Overview on {Bayesian} networks applications for dependability, risk analysis and maintenance areas.
\newblock {\em Engineering Applications of Artificial Intelligence\/}~{\em 25\/}(4), 671--682.

\bibitem[\protect\citeauthoryear{Weidinger, Schlauderer, and Overhage}{Weidinger et~al.}{2020}]{Weidinger2020}
Weidinger, J., S.~Schlauderer, and S.~Overhage (2020, December).
\newblock Information technology to the rescue? explaining the acceptance of emergency response information systems by firefighters.
\newblock {\em {IEEE} Transactions on Engineering Management\/}, 1--15.

\bibitem[\protect\citeauthoryear{Wu, Zhao, Yip, and Jiang}{Wu et~al.}{2021}]{Wu2021}
Wu, B., C.~Zhao, T.~L. Yip, and D.~Jiang (2021, March).
\newblock A novel emergency decision-making model for collision accidents in the {Yangtze} {River}.
\newblock {\em Ocean Engineering\/}~{\em 223}, 108622.

\bibitem[\protect\citeauthoryear{Wu, Shen, Wang, and Wu}{Wu et~al.}{2019}]{Wu2019}
Wu, Z., Y.~Shen, H.~Wang, and M.~Wu (2019, January).
\newblock Assessing urban flood disaster risk using {Bayesian} network model and {GIS} applications.
\newblock {\em Geomatics, Natural Hazards and Risk\/}~{\em 10\/}(1), 2163--2184.

\end{thebibliography}

\end{document}